\documentclass[a4paper,11pt]{article}	
\pdfoutput=1
\usepackage{jcappub,adjustbox}
\usepackage[T1]{fontenc}
\usepackage[utf8]{inputenc}
\usepackage[english]{babel}
\usepackage[version=4]{mhchem}
\usepackage[toc,page]{appendix}
\usepackage{lineno,mathtools,comment,graphicx,braket,booktabs,multirow,siunitx,chngcntr,vmargin,xcolor}
\usepackage{amssymb}

\newcommand{\be}{\begin{equation}}
\newcommand{\ee}{\end{equation}}

\modulolinenumbers[5]

\title{\boldmath On the relevance of prompt neutrinos for the interpretation of the IceCube signals}
\author[a]{Carlo Mascaretti,}
\author[a,b]{Francesco Vissani}

\affiliation[a]{Gran Sasso Science Institute (GSSI), Viale F. Crispi 7, 67100 L'Aquila, Italy}
\affiliation[b]{INFN, Laboratori Nazionali del Gran Sasso (LNGS), 67100 Assergi, L’Aquila, Italy}

\emailAdd{carlo.mascaretti@gssi.it}
\emailAdd{francesco.vissani@lngs.infn.it}

\abstract{
 The IceCube collaboration has discovered a new, cosmic component of high-energy neutrinos.
 Although neutrino oscillations suggest that the cosmic neutrino spectrum is almost the same for every neutrino 
 flavor, the attempts to reconstruct it, based on different analyses, lead to different energy spectra below 100 TeV. 
 In this work, we propose a phenomenological model that, assuming collisions between cosmic rays and hadrons as the production mechanism of high-energy neutrinos, yields quantitative expectations for each neutrino flavor. 
 We discuss the detectability of the prompt component of the atmospheric neutrino spectrum, pointing out the most relevant dataset, which has to be muon-neutrino depleted and to cover the energy region 10 -- 100 TeV. 
 We argue that the prompt component can cause the spectral difference between the High-Energy-Starting-Event (HESE) and the through-going muon datasets.
 Finally, we point out the need for adopting a consistent model for the interpretation of the data, stressing that a separate treatment of the different datasets is, by converse, a suboptimal procedure.
}

\begin{document}

\maketitle

\flushbottom


\section{Introduction}

After almost ten years of operation, the IceCube detector has provided unique and most important observations of neutrino events with energies ranging between \SI{100}{GeV} and \SI{10}{PeV}, which resulted in the detection of an astrophysical component of neutrinos \cite{aartsen_observation_2014,throughgoing} and in the measurement of the atmospheric components of the electron and muon neutrino spectrum \cite{ic_nue_atm,ic_numu_atm}.
The prompt component of the atmospheric neutrino spectrum, which is expected to be produced in the decay of charmed mesons in the atmosphere, has not been measured yet.

On the other hand, some missing pieces of the high-energy neutrino jigsaw puzzle seem to be finally appearing: one notable astronomical coincidence \cite{txs1} may hint that we are close to detecting the source of cosmic neutrinos.
Moreover, two double cascade events attributable to astrophysical tau neutrinos and one candidate Glashow resonance event have been recently observed\footnote{See the talk by H.~Niederhausen, on behalf of the IceCube Collaboration, at the XVIII International Workshop on Neutrino Telescopes (Venice, March 2019).}.

The accepted set of assumptions on the astrophysical component, that are adopted for the interpretation of these findings, includes: 
\begin{enumerate}
 \item isotropy
 \item standard three-flavor oscillations
 \item an unbroken power law for the energy spectrum of the new component.
\end{enumerate} 
However, the cosmic neutrino spectrum resulting from the HESE analysis \cite{ic_icrc17} is different from that obtained in the through-going muons analysis \cite{throughgoing}, as argued, for example, in \cite{palladino_icecube_2016,andrea_palladino_compatibility_2017}.
Moreover, if the spectrum from HESE were extrapolated down to energies lower than \SI{100}{TeV}, it would overshoot the gamma-ray diffuse measurements \cite{palladino_icecube_2018}.

In this situation, we find it necessary to rely on theoretical guidance.
In this work a primary cosmic-ray flux, fitted to the data of AMS-02 \cite{ams02p,ams02he} and KASCADE-Grande \cite{KG}, is defined to numerically compute both conventional and prompt components of the atmospheric neutrino spectrum.
The cosmic neutrino flux is modeled as a single-population power law, assuming $pp$ collisions in dense sources as the production mechanism. 
The ensuing expectations from theory are combined to the through-going muon analysis to make the astrophysical muon flux phenomenologically precise.
The same muon neutrino flux is also used to predict the electron and tau cosmic neutrino fluxes.
Credible regions are computed for all neutrino fluxes, which are in turn compared to the available measurements.

Finally, we comment on the results and address 
\begin{enumerate}
 \item the compelling issue of the low-energy softness of the HESE spectrum
 \item the lack of detection of prompt neutrinos in the currently examined datasets
 \item the methodological consistency of independent analyses of the various datasets.
\end{enumerate}

\section{The expected neutrino fluxes}
\label{sec:expected_fluxes}
In this section we describe our approach to the study of cosmic neutrinos, based upon theoretical considerations. 

\begin{figure}
 \includegraphics[width=\textwidth,keepaspectratio]{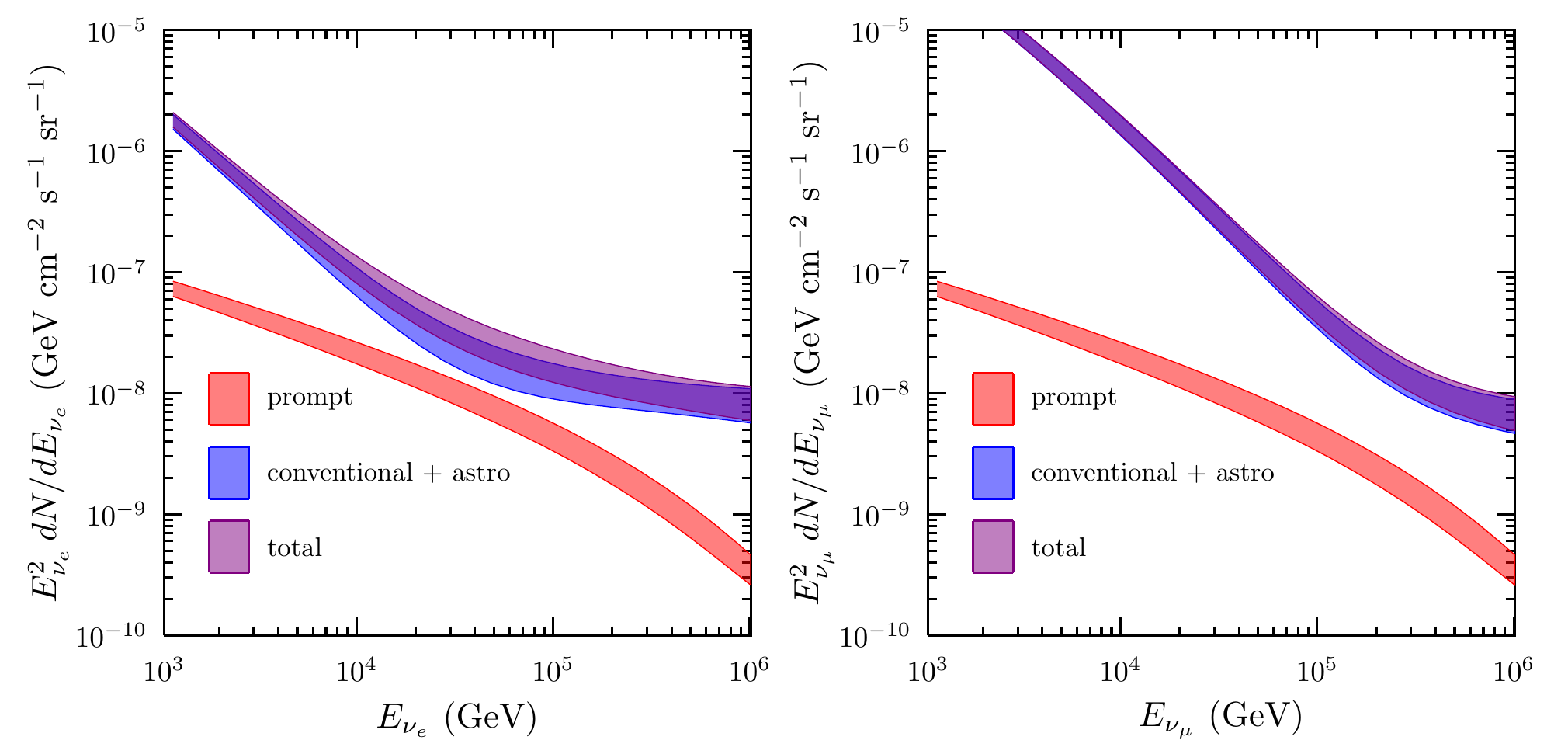}\\[5mm]
 \centering
 \includegraphics[width=0.5\textwidth,keepaspectratio]{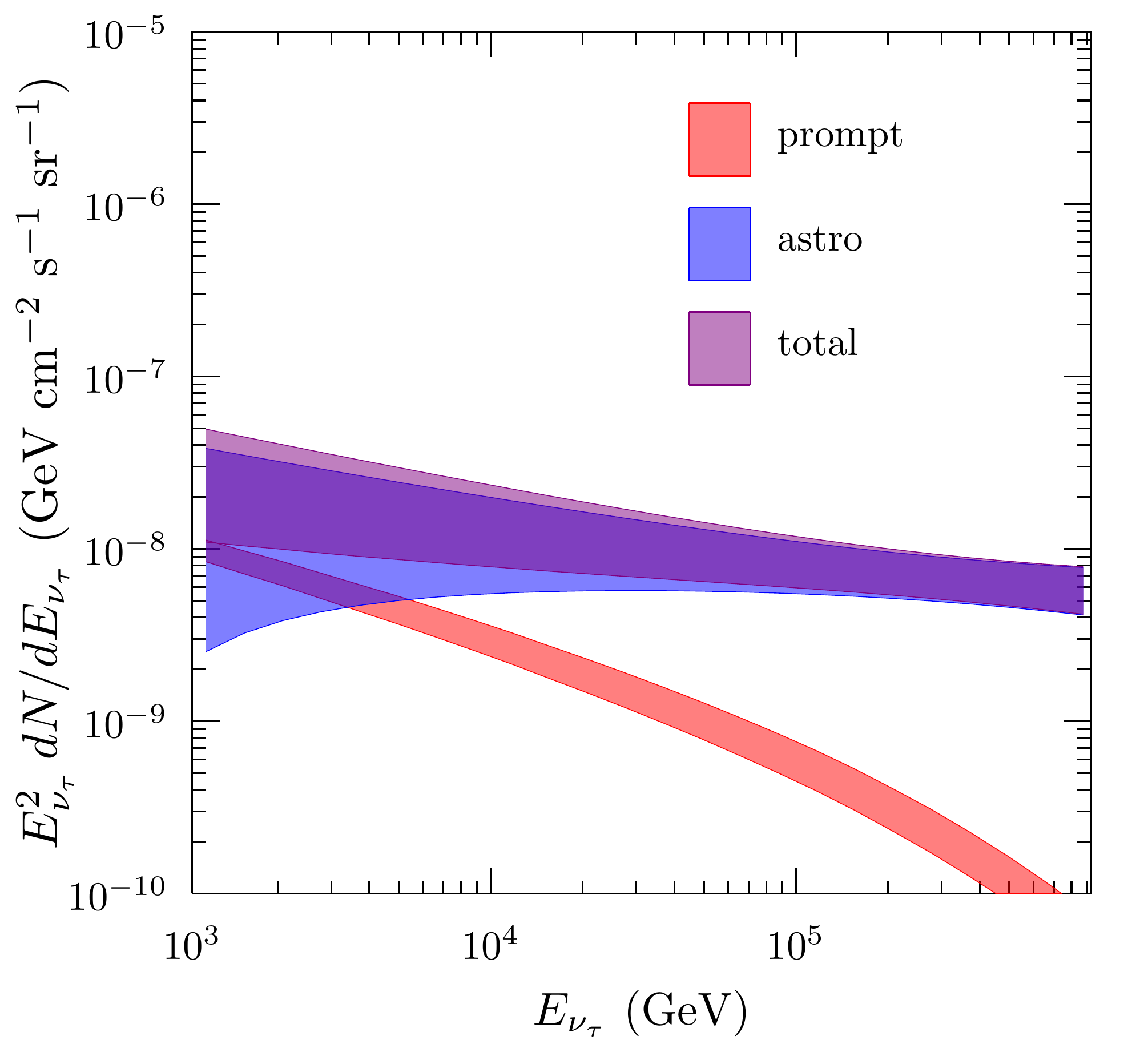}
 \caption{The expected electron (top left), muon (top right) and tau (bottom) neutrino flux components in the energy range from \SI{1}{TeV} to \SI{1}{PeV}. 
 Particular emphasis is given to prompt neutrinos, which are shown separately and summed to all other components.
 As in the rest of the paper, the bands corresponding to the sum of two or more fluxes are delimited by the sum of the lower bounds and the sum of the upper bounds of the single fluxes, as our purpose is mostly illustrative when we do not deal with single fluxes.}
 \label{fig:convastro}
\end{figure}

In figure \ref{fig:convastro} we anticipate some of our results, with the aim of emphasizing the region from \SI{1}{TeV} to \SI{1}{PeV}, where the prompt component is expected to be relevant.
This is especially evident in the panels that show the expected electron (and tau) neutrino flux components.
The normalization of the astrophysical component is an important ingredient of our model, and will be discussed in section \ref{subsec:cosmic_nus}.

\subsection{Atmospheric neutrinos}

Atmospheric neutrinos had not been studied at the highest energies before IceCube, but they can be predicted, within uncertainties linearly increasing from $\lesssim 10\%$ at \SI{100}{GeV} to about 30\% at \SI{1}{PeV}, from the observed flux of cosmic rays and the theory of strong interactions\footnote{These numbers refer to our particular computations which feature a fixed hadron interaction model, which is another source of uncertainty (around 10\% at \SI{100}{GeV} up to $30-40\%$ at \SI{100}{TeV} \cite{fedynitch_state---art_2018}).}.
Many (semi-)analytical computations \cite{TIG,sarcevic,honda,rottoli,garzelli,berss,laha_ic_2017,bhattacharya_forward_2018,giannini_intrinsic_2018} have been carried out, adopting different primary cosmic-ray and hadronic interaction models, in order to predict the {\em conventional component}, which results from the decay of pions, kaons, and unflavored mesons, and the {\em prompt component}, which comes from charmed meson decays -- even though the latter contribution is considerably more uncertain and, moreover, undetected at present. 
Before proceeding, note that atmospheric neutrinos are occasionally 
accompanied by observable muons, that are absent in the case of cosmic neutrinos instead; the observation of these coincident atmospheric muons allows us to test the reliability of the predictions for the atmospheric neutrinos, and to veto atmospheric events when looking for astrophysical ones.

There are a few good reasons to distinguish these two components: their spectrum, angular distribution, and flavor composition are different, and thus experimentally distinguishable -- in principle.

The conventional component of the atmospheric spectrum is mainly produced in the decay of pions and kaons ($\tau_\text{rest} \sim \SI{e-8}{\second}$), which are produced when cosmic rays collide with the atmosphere; since these mesons have time to interact and lose energy in the atmosphere before decaying, the spectrum of conventional atmospheric neutrinos is expected to be softer than that of the parent cosmic rays of a factor $E^{-1}$, i.e.~$\propto E^{-3.7}$.
Prompt neutrinos are also produced in hadronic collisions of cosmic rays on the atmosphere as a product of the subsequent immediate decay of charmed mesons ($\tau_\text{rest} \sim \SI{e-12}{\second}$), so that their spectrum is expected to closely reflect that of the parent cosmic rays $\propto E^{-2.7}$.

The bulk of atmospheric neutrinos is due to pions and kaons; the larger the energies of the muon produced in the pion/kaon decay, the smaller the probability it decays, so that the flavor ratio of such component shifts from a $(1:2:0)$ for energies lower than \SI{10}{GeV} to $(0:1:0)$ at higher energies.
The conventional component is also anisotropic: the thicker the layer of the atmosphere traversed (with maximum at an azimuthal angle $\theta = \pi/2$, where $\theta=0$ refers to the upward direction and $\theta = \pi$ to the downward one), the larger the amount of possible targets for cosmic rays.

The prompt component is expected, on the contrary, to be isotropic and to follow a flavor ratio of about $(1:1:0.1)$, with the same energy distribution for all flavors\footnote{{Expectations based on MCEq \cite{MCEQ} and the branching ratios of the relevant mesons \cite{tanabashi_review_2018}.}}. 
This means that the prompt contribution is the dominant one in the $\nu_e$ atmospheric spectrum already at few tens TeV, as we will later show, while, in the case of $\nu_\mu$, it reaches the level of the conventional one only at larger-than-PeV energies.

Following \cite{mbe_2019}, we compute the conventional and prompt atmospheric neutrino flux with MCEq \cite{MCEQ}, adopting the most recent version of SYBILL, the 2.3c release \cite{syb2.3}, the NRLMSISE-00 \cite{nrlmsise} model of the atmosphere, and a primary CR flux defined as follows: 
\begin{itemize}
 \item only protons and helium nuclei are considered, because they are the most abundant elemental species, and because nuclei of mass number $A$ and energy $E$ produce neutrinos with average energy $E/(20A)$ when colliding with other nuclei;
 \item the most important part of the (Galactic) CR flux is a power-law fitted to the AMS-02 \cite{ams02p,ams02he} low energy (from 100 GeV to 10 TeV) data, and its knee is assumed to be either an ``exponential-square'' cutoff or a change of slope. 
 \begin{equation}
 \frac{d\Phi_{p,\text{He}}^\text{exp2-cut}}{dE}=N_{p,\text{He}}\left(\frac{E}{\SI{10}{TeV}}\right)^{-\gamma_{p,\text{He}}} \exp\left[-\left(\frac{E}{Z_{p,\text{He}}R_\text{knee}}\right)^2\right] \label{eq:primaries_exp2} 
 \end{equation}
 \begin{equation}
 \frac{d\Phi_{p,\text{He}}^\text{delta-slope}}{dE}=N_{p,\text{He}}\left(\frac{E}{\SI{10}{TeV}}\right)^{-\gamma_{p,\text{He}}} \times \begin{dcases} 1 & E\leq Z_{p,\text{He}}R_\text{knee} \\ \left(\frac{E}{\SI{10}{TeV}} \right)^{-\alpha} & E > Z_{p,\text{He}}R_\text{knee}
 \end{dcases} \label{eq:primaries_dslope} 
 \end{equation}
 where $\alpha = 2-\delta$, $\delta =1/3 \equiv$ slope of the diffusion coefficient, i.e.~$D(E) = D_0 (E/E_0)^\delta$.
 The knee is assumed to be rigidity-dependent, and its position is obtained by fitting the overall primary shape of eq.~\eqref{eq:primaries} to the KASCADE-Grande data \cite{KG};
 \item an additional, supposedly extra-galactic\footnote{In \cite{KG} this is the component which onsets after the ankle at $E \approx \SI{e17}{\electronvolt}$.} proton component is added to the fitting spectrum
 \begin{equation}
 \frac{d \Phi_\text{eg}}{d E }= N_\text{eg}\left(\frac{E}{\SI{100}{PeV}} \right)^{-2.7} 
 \label{eq:egp}
 \end{equation}
 with $N_\text{eg}$ as the only free parameter.
 
 \begin{equation}
 \frac{d \Phi_\text{tot}^k}{d E}=\sum_{i=p,\text{He}}\frac{d \Phi_i^k}{d E}+\frac{d \Phi_\text{eg}}{d E}
 \label{eq:primaries} 
 \end{equation}
where the index $k$ identifies the hypothesis on the knee of cosmic rays: its value is either $k={\tt exp2}${\tt-cut} for an exponential-square cutoff or  $k ={\tt delta}${\tt-slope} for a change of slope.
\end{itemize}
The errors on the atmospheric neutrino flux are given by the uncertainty on the shape of the knee and by that on the normalisation and slope resulting from the fits.
In table \ref{tab:fitresults} we show the results of our fits.

\begin{table}[htp]
 \centering
\begin{adjustbox}{max width=\textwidth}
 \begin{tabular}{lcccccc}
 \midrule
Model &$R_\text{knee}$ &$N_p$ &$\gamma_p$ &$N_\text{He}$ &$\gamma_\text{He}$ &$N_\text{eg}$ \\
 \midrule
 {\tt exp2-cut} &$15.1 \pm \SI{0.7}{PV}$ &\multirow{2}{*}{$1.5\pm 0.2$} &\multirow{2}{*}{$2.71\pm 0.04$} &\multirow{2}{*}{$1.5\pm 0.1$} &\multirow{2}{*}{$2.64\pm 0.03$} 
 & $6.0 \pm 0.2$
 \\ 
{\tt delta-slope}& $5.8 \pm \SI{0.6}{PV} $ & & & && $5.0 \pm 0.5 $ \\
 \midrule
 \end{tabular}
 \end{adjustbox}
 \caption{The parameters of our primary CR spectrum as resulting from the fit to the KASCADE-Grande data with the two knee models of eq.~\eqref{eq:primaries_exp2} and \eqref{eq:primaries_dslope}. $N_p$ and $N_\text{He}$ are given in units of \SI{e-7}{\per\giga\electronvolt\per\square\meter\per\second\per\steradian}, while $N_\text{eg}$ is in units of \SI{e-19}{\per\giga\electronvolt\per\square\meter\per\second\per\steradian}. 
 See \cite{mbe_2019} for more details on this.}
 \label{tab:fitresults}
\end{table}

\subsection{Cosmic neutrinos}
\label{subsec:cosmic_nus}

The predictions for cosmic neutrinos are much more uncertain and require more discussion. Depending upon the adopted model, the resulting neutrino spectra vary greatly in shape and normalization. 
One of the few stable expectations is that, according to the observed three flavor oscillation phenomena, cosmic electron, muon and tau neutrinos have to be present in similar amounts. 
The simplest and most popular hypothesis is that cosmic neutrinos are distributed as $E_\nu^{-\gamma}$ with $\gamma\sim 2$, at least in the range of energies where they become observable. Surely this case is conducive to observation, however it is important to state clearly what are its motivations, what is its extent, what are its implications; this is the aim of the present discussion.
 
The physical picture that we have in mind is that cosmic neutrinos are produced in collisions between the accelerated cosmic rays and the gas surrounding the accelerators. 
In the diffusive shock acceleration (DSA) picture, the cosmic-ray spectrum is a power-law $\propto E^{-2}$; due to the scaling associated to hadronic collisions, also the gamma-ray and neutrino spectra at the source\footnote{The neutrino spectrum is supposed to be unaltered even far from the source.} will be power laws $\propto E^{-2}$.
This setup can be regarded as an extension of what it is commonly supposed for the Galactic cosmic rays, where $E_\text{max}$ has probably a lower value than $10-\SI{100}{PeV}$; however, the slope of the injected cosmic rays might be similar or equal to the Galactic one. 
The abundance of target hadrons points out instead to some dusty environment; this could the site of intense stellar formation, say, starburst and/or star-forming Galaxies. 

As a specific instance, we refer to the theoretical model of Loeb and Waxman \cite{waxman}, according to whom the flux of cosmic neutrinos is:
\begin{equation}
 \frac{d\Phi}{d E} = \Phi_\text{astro} \left(\frac{E}{\SI{100}{TeV}} \right)^{-\gamma_\text{astro}}
 \label{eq:waxman-loeb}
\end{equation}
with 
\[\Phi_\text{astro}^\text{LW} = 2 \times 10^{\pm 0.5} \times \SI{e-18}{\per\giga\electronvolt\per\square\centi\meter\per\second\per\steradian} \qquad \gamma_\text{astro}^\text{LW} = 2.15\pm 0.10\]
in the case (as we know today) of an astrophysical neutrino spectrum extending above \SI{100}{TeV}.
The slope of this spectrum is very close to the one we would expect in the simple case of DSA. 

The assumptions above are consistent with the measurements of through-going muons \cite{throughgoing} obtained by IceCube above \SI{200}{TeV} and tested by the HESE dataset \cite{aartsen_observation_2014} in the same energy region, which result in 
\[ \Phi_\text{astro}^{\text{IC},\mu} = 0.90^{+0.30}_{-0.27}\times \SI{e-18}{\per\giga\electronvolt\per\square\centi\meter\per\second\per\steradian} \qquad \gamma_\text{astro}^{\text{IC},\mu} = 2.13\pm 0.13\]
Moreover, as we will show in the following, they do not imply any clash with the measurements of the diffuse gamma-ray emission, obtained at lower energies, even if we assume that the cosmic neutrino spectrum extends well below the observed range. 

{
Before proceeding, let us summarize our position. 
Our goal is to explore the hypothesis that the signal of cosmic neutrinos be produced mainly in $pp$ collisions, a hypothesis consistent with IceCube's observations at the highest energies.
Admittedly, there is no strong theoretical reason as of now to exclude a concurrent, or even dominant, $p\gamma$ origin; thus, we admit that our primary motivation for exploring the $pp$ hypothesis first is based on the Occam razor.
On the other hand, the attempts of proceeding beyond this stage of the discussion, by performing accurate modelling of the astrophysical signal, have just begun and do not have yet the compelling and quantitative character that have, say, the modelling of atmospheric neutrinos discussed above.
}
\bigskip

Assuming $pp$-based sources of neutrinos, we expect pion decay as the main mechanism of neutrino production, thus resulting in a $(\nu_e:\nu_\mu:\nu_\tau) \simeq (1:2:0)$ flavor ratio at the source; due to the impact of standard three-flavor neutrino oscillations, we expect a flavor ratio at Earth of about $(1:1:1)$.

In order to have much more precise predictions, we define our phenomenological muon neutrino flux by combining the through-going muon neutrino flux as observed in \cite{throughgoing} and that of Loeb and Waxman of eq.~\eqref{eq:waxman-loeb}.
We do this in the following way:
\begin{enumerate}
 \item we define the combined $\gamma_\text{astro}$ and $\Phi_\text{astro}$ - as well as their errors - as the weighted average of those from \cite{waxman} and \cite{throughgoing}; notice that, due to the large error on the normalization from \cite{waxman}, the weighted average on the normalization is very close to that of \cite{throughgoing}, while $\gamma_\text{astro}^\text{best} = 2.14 \pm 0.08$. 
 The resulting combined muon neutrino flux is thus:
 \begin{equation}
 \frac{d\Phi_{\nu_\mu}}{d E} = 0.90^{+0.30}_{-0.27} \times 10^{-18} \left(\frac{E}{\SI{100}{TeV}} \right)^{-2.14\pm 0.08}\;\si{\per\giga\electronvolt\per\square\centi\meter\per\second\per\steradian}
 \label{eq:astronu_flux}
 \end{equation}
 
 \item we reproduce, using a Gaussian likelihood, the 68\% CL contour in the $\gamma_\text{astro}$-$\Phi_\text{astro}$ plane in figure 6 of \cite{throughgoing} in order to account for the correlation\footnote{Shown in figure 3 of the same paper.} $\rho \sim 0.6$ between the two parameters:
 \[\mathcal L(\mathbf v)=\frac{1}{2\pi\sqrt{\det\Sigma^2}} \exp\left[-\frac{1}{2}(\mathbf v -\mathbf v^\text{best})^T \Sigma^{-2} (\mathbf v -\mathbf v^\text{best}) \right] \] 
 where
 \[\mathbf v = \begin{pmatrix} \Phi_\text{astro} \\ \gamma_\text{astro} \end{pmatrix} \qquad \Sigma^2 = \begin{pmatrix} \sigma_{\Phi}^2 & \rho \sigma_{\Phi}\sigma_{\gamma} \\ \rho \sigma_{\Phi}\sigma_{\gamma} & \sigma_\gamma^2 \end{pmatrix} \]
 and $\sigma_\Phi$ and $\sigma_\gamma$ are the errors on the flux normalization at \SI{100}{TeV} and the slope respectively, as taken from \cite{throughgoing}, since the correlation of the parameters can be extracted only from the data;
 
 \item we define the best fit astrophysical neutrino flux as the flux averaged using the likelihood defined above:
 \be
 \Braket{\frac{d\Phi}{dE}} = \int_{\Phi_\text{astro}^\text{best}-5\sigma_{\Phi}}^{\Phi_\text{astro}^\text{best}+5\sigma_{\Phi}}d\Phi_\text{astro} \int_{\gamma_\text{astro}^\text{best}-5\sigma_{\gamma}}^{\gamma_\text{astro}^\text{best}+5\sigma_{\gamma}}d\gamma_\text{astro} \,\mathcal L(\Phi_\text{astro},\gamma_\text{astro})\,
 \frac{d\Phi}{dE}
 \label{eq:av_flux}
 \ee
 and its 1$\sigma$ uncertainty as:
 \be 
 \delta\left(\frac{d\Phi}{dE} \right)=\sqrt{\Braket{\left(\frac{d\Phi}{dE}\right)^2}-\left(\Braket{\frac{d\Phi}{dE}}\right)^2} 
 \label{eq:std_flux}
 \ee
\end{enumerate}
The resulting cosmic muon neutrino flux is the blue band in figure \ref{fig:numu_components}.
We can obtain the cosmic flux of $\nu_e$ and $\nu_\tau$ simply by multiplying that of muon neutrinos by $R_{e\mu}$ and $R_{\tau\mu}$
\[\frac{d\Phi_{\nu_e}}{d E}=R_{e\mu} \frac{d\Phi_{\nu_\mu}}{d E} \qquad \frac{d\Phi_{\nu_\tau}}{d E}=R_{\tau\mu} \frac{d\Phi_{\nu_\mu}}{d E} \]
In fact, the $R_{\ell\mu}$ factors can be calculated in \emph{two different approximations}. Standard three-flavor oscillations are assumed in both cases, and described by $P_{\ell\ell'}$, the matrix of the survival/oscillation probabilities averaged over cosmic distances
\[P_{\ell\ell'} = \sum_{i=1}^3|U_{\ell i}^2||U_{\ell'i}^2| \]
where $U$ is the neutrino mixing matrix. 
The two approaches are the following:
\begin{enumerate}
 \item the ``$2:1$ approximation'', in which we define
 \[\displaystyle R_{\ell\ell'} = \frac{\sum_{\ell''} P_{\ell\ell''}\xi_{\ell''}^0}{\sum_{\ell''} P_{\ell'\ell''}\xi^0_{\ell''}} \qquad \ell,\ell',\ell'' = e,\mu,\tau \]
 where $\xi_\ell^0$ is the fraction of $\nu_\ell+\overline \nu_\ell$ produced at the source.
 We compute $R_{\ell\ell'}$ assuming the commonly accepted flavor ratio\footnote{Known as the pion decay scenario. Choosing a generic flavor ratio with $\xi_\mu^0 = x$, $\xi_e^0=1-x$, $\xi_\tau^0 = 0$, sampling $x$ uniformly between 0 and 1, we would have $R_{e\mu}= 0.78^{+0.57}_{-0.07}$ and $R_{\tau\mu}=1.00^{+0.05}_{-0.15}$.} at production of $\xi_\mu^0=2/3$ and $\xi_\tau^0 = 0$ and sampling $P_{\ell\ell'}$ according to their distributions (see Appendix \ref{appendix2}) in the case of normal hierarchy of the neutrino masses.
 We obtain:
 \[R_{e\mu} = 1.09^{+0.03}_{-0.04} \qquad R_{\tau\mu} = 0.97^{+0.03}_{-0.04} \]
 
 \item the ``kernel approach'', which relies on a more accurate and physically more comprehensive procedure to fully embrace the consequences of the hadronic production mechanism, namely the strict relationship between gamma rays and neutrinos.
 As shown in \cite{villaviss}, and updated in Appendix \ref{appendix2} with the use of the most recent oscillation parameters from \cite{bari}, the gamma-ray flux at the production site and the cosmic neutrino flux are linked by the following relation:
 \begin{equation}\frac{d\Phi_{\nu}(E_\nu)}{dE_\nu } = \int_{E_\nu}^{\infty} \frac{dE}{E}\,\tilde K_\nu(E_\nu,E)\,\frac{d\Phi_\gamma(E)}{dE} 
 \label{eq:eq5}
 \end{equation}
 where $\tilde K_\nu$ is a kernel which accounts also for $\nu$ oscillations (see \cite{villaviss} and Appendix \ref{appendix1} for more on this).
 Equation ~\eqref{eq:eq5} can be rewritten as
 \begin{equation} \frac{d\Phi_{\nu}(E_\nu)}{dE_\nu } = \int_{0}^{1} \frac{dx}{x}\,\tilde K(x)\,\frac{d\Phi_\gamma(x/E_\nu)}{dE} 
  \label{eq:appendila}
  \end{equation}
 and it can be easily proven that, assuming a power-law gamma-ray flux $\propto E^{-\gamma}$:
 \begin{equation}
 R_{\ell\ell'}=\frac{\zeta_{\nu_\ell}(\gamma)}{\zeta_{\nu_{\ell'}}(\gamma)}
 \label{eq:zfactors_ratio}
 \end{equation}
 where 
 \[ \zeta_{\nu_\ell}(\gamma) = \int_0^1 dx\,x^{\gamma-1} \left[\tilde K_{\nu_\ell}(x)+\tilde K_{\overline\nu_\ell}(x) \right] \]
 $R_{\ell\ell'}$ as defined in eq.~\eqref{eq:zfactors_ratio} depend negligibly\footnote{They change of less than $0.1\%$ in the $3\sigma$ range around $\gamma_\text{astro}^\text{best}=2.14$.} on $\gamma$, and this time we obtained:
 \[R_{e\mu}= 1.30 \pm 0.05 \qquad R_{\tau\mu} =0.92 \pm 0.04\]
 We conservatively estimated the impact a 20\% variation of the non-oscillated electron (anti-)neutrino kernel variation, which could be due to systematic errors, as neglecting the $K_L$ and $K_S$ contributions \cite{villaviss}: we found out that $R_{e\mu}$ varies of 6\%, while $R_{\tau\mu}$ of 1\%, which are not important considering the $30\%$ uncertainty on the astrophysical muon neutrino flux normalization.
 Moreover, if the electron neutrino and muon neutrino kernels were subject to the same systematic error, the factors $R_{\ell\ell'}$ would not change.
\end{enumerate}
Both procedures yield very similar tau neutrino fluxes, while the electron neutrino flux is larger in the kernel approach than in the $2:1$ approximation.
This is due to the fact that in the kernel formalism the presence of neutrinos due to kaons is accounted for, the decay of which results in an electron-neutrino richer flux.

We chose to multiply the flux of muon neutrinos by $R_{\ell\ell'}$ computed with the kernel approach to consider hadronic collisions as the neutrino production mechanism and include the effect of charged kaons, having thus a clear and precise physical picture in mind.
Note that the error on the astrophysical $\nu_\mu$ flux normalization is 30\%, much larger that those on $R_{e\mu}$ and $R_{\tau\mu}$, so that the relative error on the cosmic electron and tau neutrino fluxes will be 30\% as well.

\section{Results and discussion}

\subsection{The components of the neutrino spectra}

\begin{figure}
 \centering
 \includegraphics[width=\textwidth,keepaspectratio]{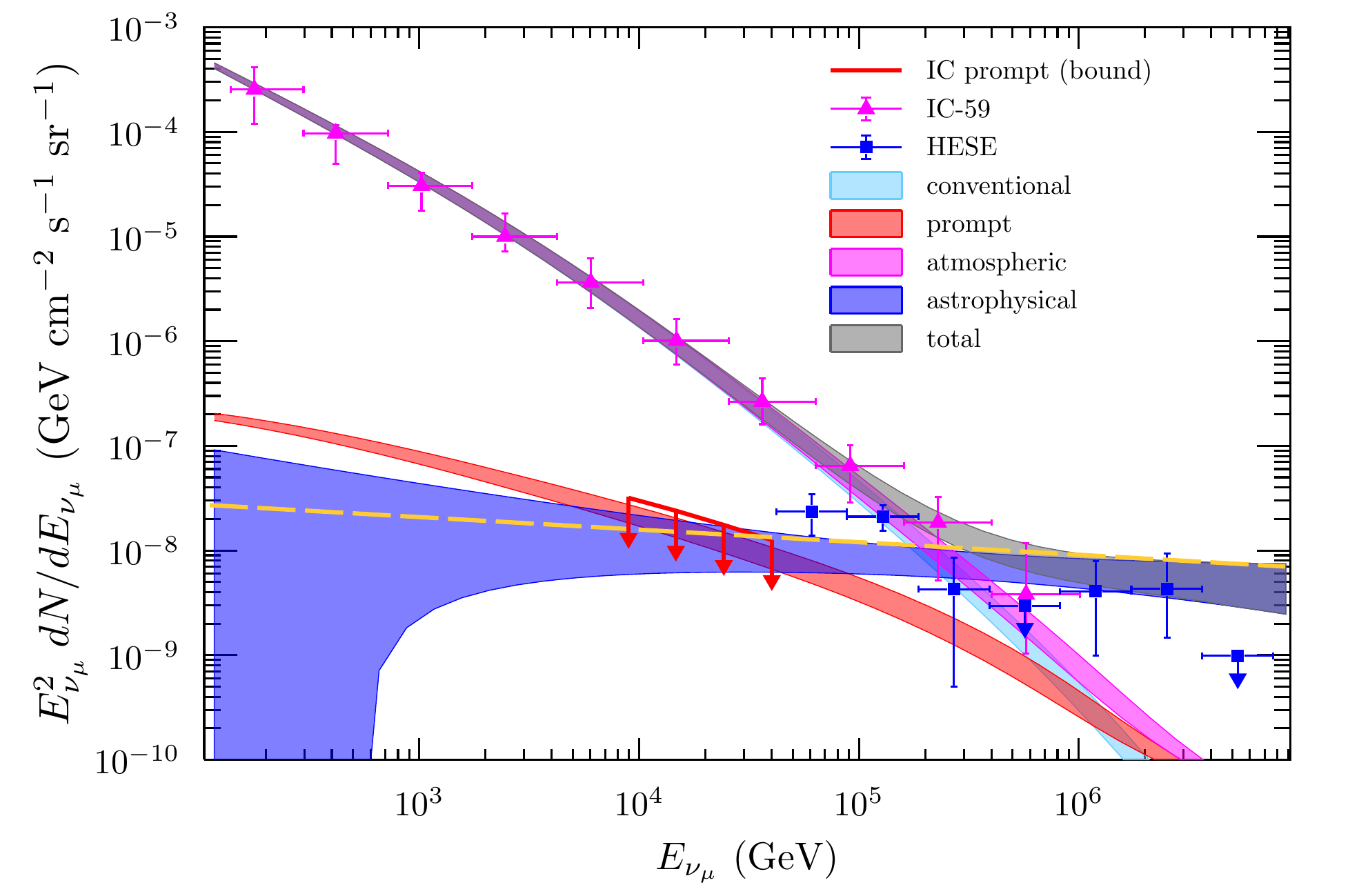}
 \caption{The various components of the muon neutrino flux obtained with the model defined in Section \ref{sec:expected_fluxes}. Also shown are the measurement of the atmospheric muon neutrino flux by IceCube \cite{ic_numu_atm}, that of the cosmic neutrino flux with from the HESE dataset \cite{ic_icrc17}, the 68\% C.L.~upper bound on prompt muon neutrinos in the relevant sensitivity region \cite{throughgoing} (red line with arrows) and the upper flux limit (yellow dashed line) obtained in \cite{palladino_icecube_2018}, featuring $\gamma=2.12$ and the best fit normalization + 1$\sigma$ as taken from the through-going muon analysis \cite{throughgoing}.}
 \label{fig:numu_components}
\end{figure}

\begin{figure}
 \centering
 \includegraphics[width=\textwidth,keepaspectratio]{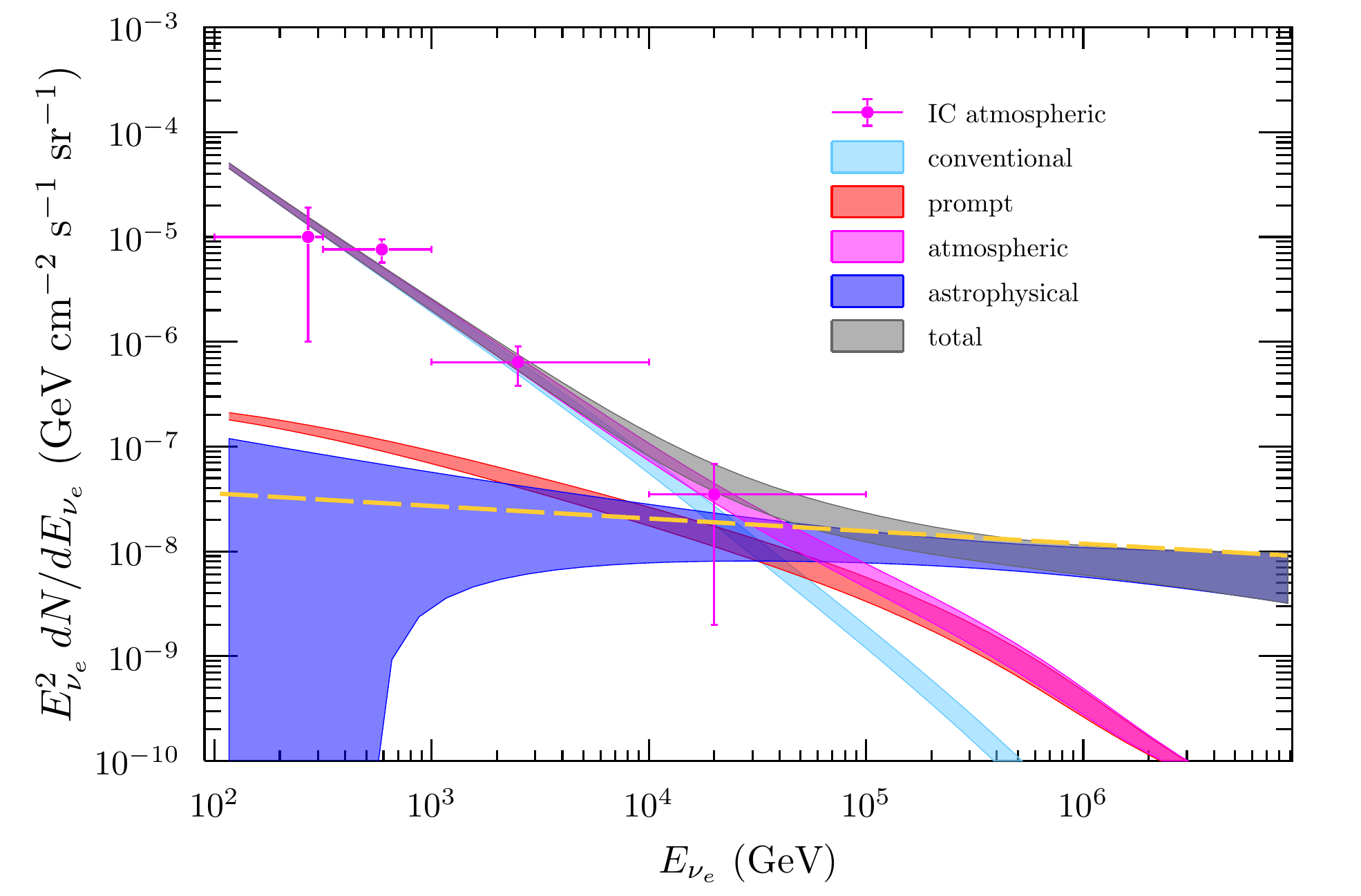}
 \caption{The predictions of the various components of the electron neutrino flux obtained with the model defined in Section \ref{sec:expected_fluxes}. Also shown is the measurement of the atmospheric electron neutrino flux by IceCube \cite{ic_nue_atm} and the upper flux limit (yellow dashed line) obtained in \cite{palladino_icecube_2018}, featuring $\gamma=2.12$ and the best fit normalization + 1$\sigma$ as taken from the through-going muon analysis from \cite{throughgoing} rescaled by $\zeta_{\nu_e}/\zeta_{\nu_\mu}$.
 }
 \label{fig:nue_components}
\end{figure}

In figure \ref{fig:numu_components} we show the expectations for the various components of the muon neutrino flux, as well as the corresponding measurements by the IceCube Collaboration.
We assume isotropy, but note that the IC-59 points and the HESE ones refer to two different kinds of events, namely $\nu_\mu$-induced tracks from the Northern sky and all-flavor High-Energy Starting Events from the whole sky respectively.

In figure \ref{fig:nue_components} we show the prediction for the cosmic electron neutrino flux and the expectations for the other components of the electron neutrino spectrum, alongside with the relevant measurement by the IceCube Collaboration.

No tau neutrino data is available, also due to the fact that the atmospheric tau neutrino flux consists only in the prompt component, thus we do not show the corresponding plot as it would convey no useful information.

From figure \ref{fig:numu_components} and \ref{fig:nue_components} a few noteworthy features are noticeable:
\begin{enumerate}
 \item the conventional atmospheric expectations obtained with MCEq, featuring SYBILL-2.3c and our primary cosmic ray spectrum eq.~\eqref{eq:primaries}, agree well with the measurements from IceCube;
 \item the region where the atmospheric and cosmic components of the muon neutrino spectrum cross is around $\SI{250}{TeV} \simeq E_\text{knee}/20$;
 \item the prompt component is always subdominant in the $\nu_\mu$ spectrum, so that within our model it is not surprising that no significant evidence for prompt neutrinos has been found in the through-going muons analysis;
 \item the conventional atmospheric component is a factor $\sim 30$ less important for electron neutrinos compared to muon neutrinos, so that the prompt component sizably contributes to the overall flux of $\nu_e$ for $E_\nu \geq \SI{10}{TeV}$.

\end{enumerate}

\subsection{Is it possible to extract the prompt neutrino signal?}

Our results suggest that the best chance to detect the prompt flux of neutrinos is from electron neutrino atmospheric data for $E\gtrsim \SI{1}{TeV}$, assuming the possibility to somehow discriminate the flavor of the events.

The cascade (or shower) event topology is the most interesting in this regard; it is one of the two kinds of events comprised in the HESE dataset, the remainder being tracks.
Track-like events are produced by charged-current muon neutrino interactions, while cascades are produced in all other possible cases.
It follows that the cascade sample is the one with the smallest relative contribution of muon neutrinos, which, as already seen in figure \ref{fig:numu_components}, are very prompt-neutrino poor. 
Another reason to focus on the cascade subset of the HESE dataset is that the track-like subset is compatible with being due to background events (atmospheric muons and atmospheric muon neutrinos) only \cite{andrea_palladino_compatibility_2017,niederhausen_high_2018}.

The contributors to the cascade dataset are: 
\begin{enumerate}
 \item atmospheric muons;
 \item conventional atmospheric $\nu_e$ and $\nu_\mu$;
 \item prompt atmospheric $\nu_e$, $\nu_\mu$ and $\nu_\tau$;
 \item astrophysical $\nu_e$, $\nu_\mu$ and $\nu_\tau$.
\end{enumerate}
Therefore, provided that the contamination due to muons, conventional muon neutrinos and all-flavor astrophysical neutrinos can be subtracted, the cascade sample offers us the chance to detect prompt neutrinos.

\begin{figure}
 \centering
 \includegraphics[width=.46\textwidth]{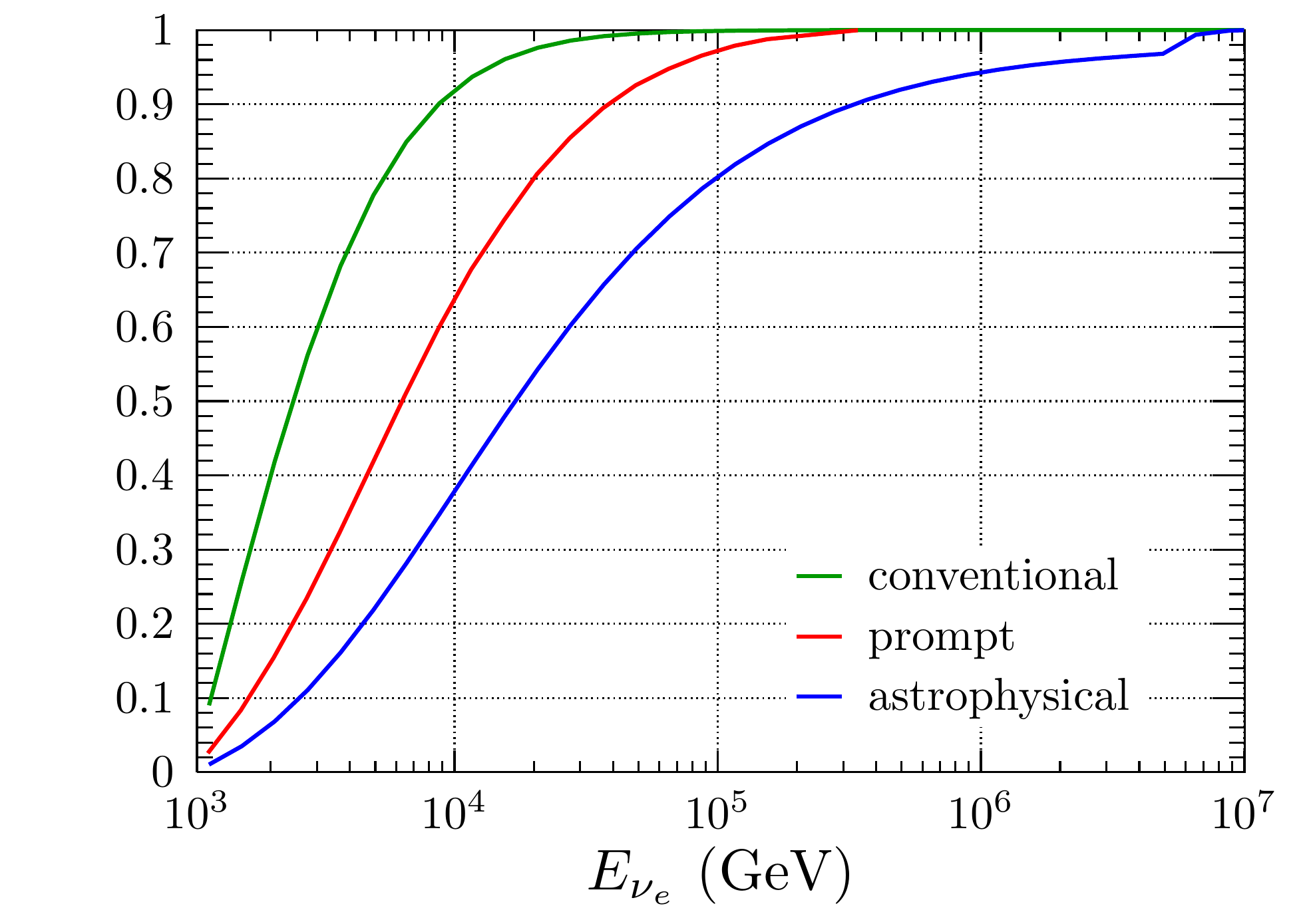} \hfill
 \includegraphics[width=.46\textwidth]{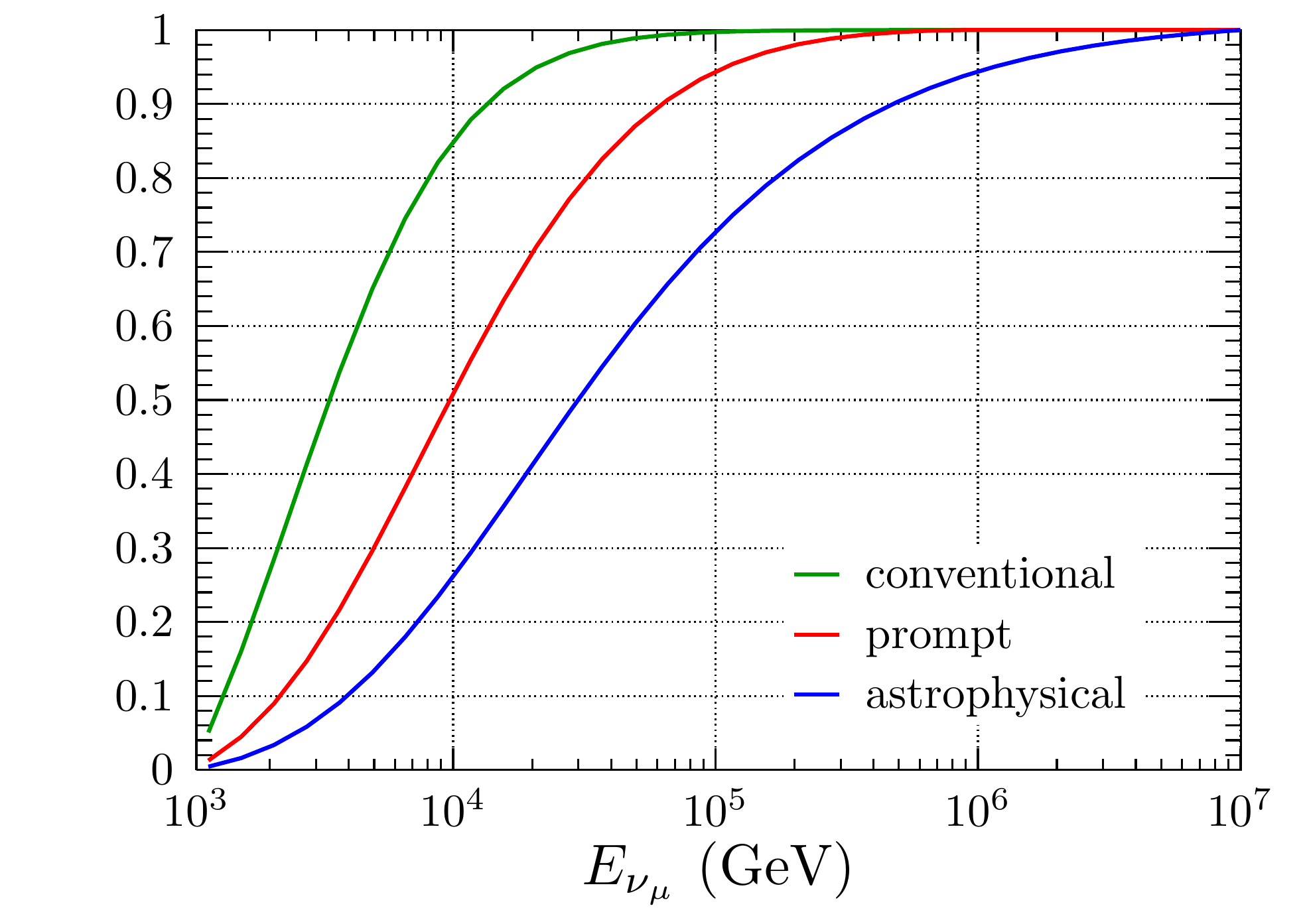}\\ \vfill
 \centering
 \includegraphics[width=.46\textwidth]{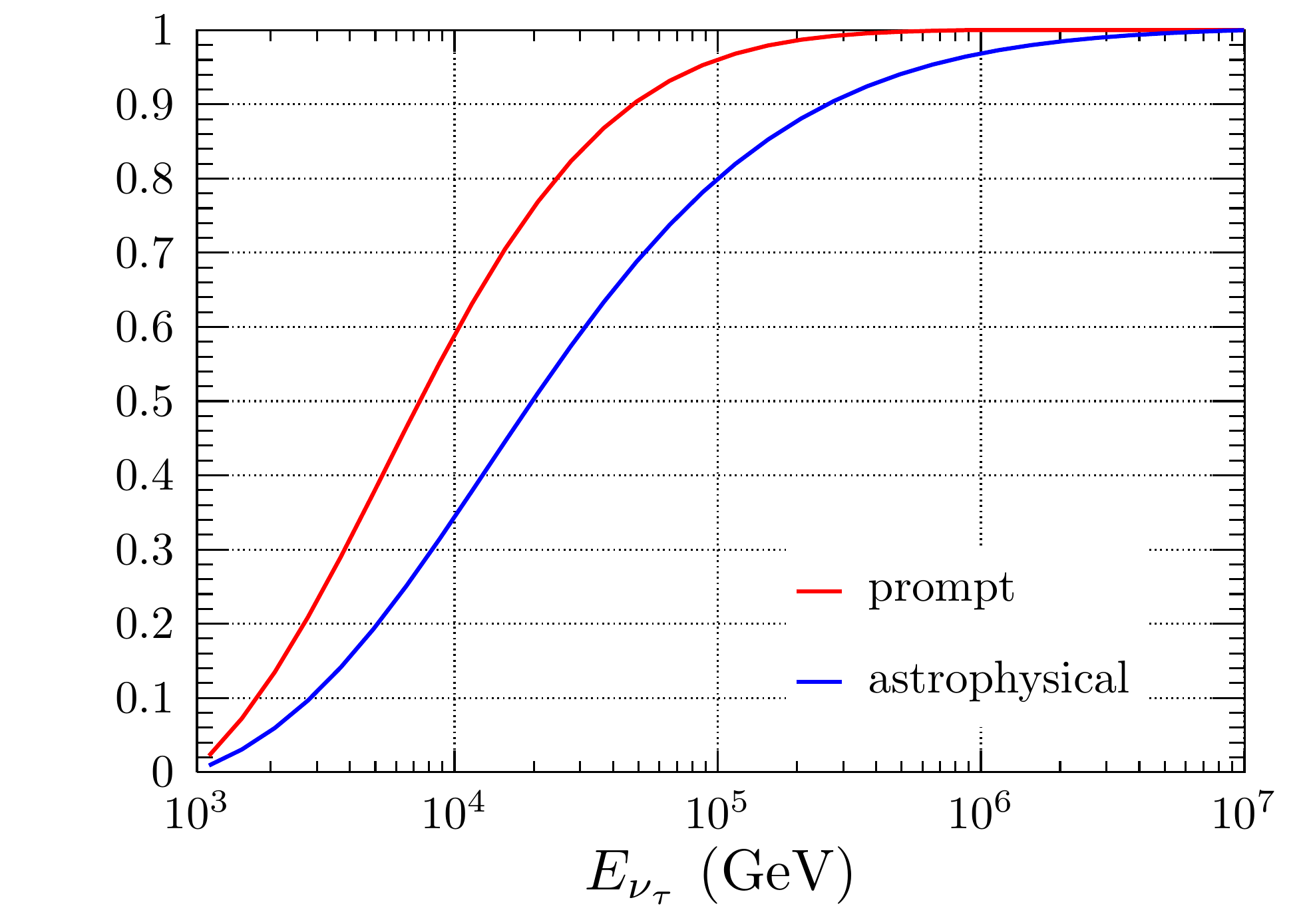}
 \caption{The cumulative distributions of cascade events in IceCube in the energy range \SI{1}{TeV} -- \SI{10}{PeV}, divided by component and flavor.}
 \label{fig:parent}
\end{figure}

In order to quantitatively test our hypothesis we use the effective areas for cascade-like events given in figure 1 of \cite{niederhausen_high_2018} to estimate the yearly rate of cascade-like events due to neutrinos with larger-than-TeV energy.
In figure \ref{fig:parent} we show the parent distribution of cascade events, dividing them by flavor and component, in order to show the energy ranges which contribute the most to the events of table \ref{tab:showers}.
The yearly rates are computed according to:
\be 
\Gamma_{\nu_\ell} = 4\pi\times \SI{1}{year} \times \int_{\SI{1}{TeV}}^{\SI{10}{PeV}} dE\,\mathcal A_{\nu_\ell}(E) \,\frac{d\Phi_{\nu_\ell}}{d E} 
\label{eq:showers}
\ee
where $\mathcal A_{\nu_\ell}(E)$ is the effective area for the detection of cascade-like events induced by $\nu_\ell$ and $\overline \nu_\ell$, and $\SI{1}{year}=\SI{\pi e7}{s}$.

The expected rate due to prompt neutrinos is small, less than 10\% of the conventional contribution, which is not encouraging for the search of prompt neutrinos.
However, the conventional contribution could be somewhat reduced by searching for cascades coming from below and/or by using a higher energy threshold, so as to exclude most of the conventional events.
 
\vfill
\begin{table}[htbp]
 \centering
 \begin{tabular}{lccccc}
 \midrule
 Component &&$\Gamma_{\nu_e}$ &$\Gamma_{\nu_\mu}$ &$\Gamma_{\nu_\tau}$ &$\Gamma_\text{tot}$ \\ \midrule
 Conventional && 160 -- 210 &420 -- 570 &0 &580 -- 780\\ \midrule
 Prompt && 20 -- 30 &3 -- 5 &2 -- 3 &25 -- 40\\ \midrule
 Cosmic && 10 -- 40 & 2 -- 6 & 5 -- 20 &15 -- 65\\ \midrule
 \end{tabular}
 \caption{$68\%$ CL intervals relative to the yearly rate of cascade-like events in the energy range \SI{1}{TeV} and \SI{10}{PeV} in IceCube, as computed with eq.~\eqref{eq:showers}, due to the different components of the three neutrino fluxes.}
 \label{tab:showers}
\end{table}
\subsection{Is there a spectral anomaly?}

In \cite{palladino_icecube_2016,andrea_palladino_compatibility_2017,denton_invisible_2018} the spectral difference between the HESE and the through-going muons spectra is labelled as an anomaly, and it has been already addressed in \cite{palladino_multi-component_2018,sui_combined_2018}.
In fact, if we compare the astrophysical neutrino fits to the cascade and starting track samples, which together constitute the HESE dataset, from \cite{niederhausen_high_2018} to that from the through-going muons analysis \cite{throughgoing} there is an evident difference, as can be appreciated from table \ref{tab:spectral_differences} and from figure 10 of \cite{niederhausen_high_2018}.

\begin{table}[htbp]
 \centering
 \begin{tabular}{lcccc}
 \midrule
 dataset &&$\Phi_\text{astro}$ &&$\gamma_\text{astro}$ \\ \midrule
 C && $2.2^{+0.6}_{-0.5}$ && $2.62 \pm 0.08$\\ \midrule
 ST && $1.6^{+1.6}_{-1.0}$ && $2.43^{+0.28}_{-0.30}$\\ \midrule
 HESE && $2.46 \pm 0.8$ && $2.92^{+0.33}_{-0.29}$\\ \midrule
 TM && ${0.90}^{+0.30}_{-0.27} $ && $2.13 \pm 0.13 $ \\ \midrule
 \end{tabular}
 \caption{The flux normalizations, in units of \SI{e-18}{\per\giga\electronvolt\per\square\centi\meter\per\second\per\steradian}, and slopes deriving from the astrophysical best fits to the cascade (C) and starting tracks (ST) samples from \cite{niederhausen_high_2018}, to the 6-years HESE sample from \cite{kopper_observation_2018}  and to the through-going muons (TM) sample from \cite{throughgoing}. 
 The numerical values of the flux normalizations for the cascade and starting tracks best fits have been obtained from figure 10 of \cite{niederhausen_high_2018} as they are not explicitly reported in the paper.}
 \label{tab:spectral_differences}
\end{table}

Since, as said before, the starting track sample is compatible to be due to background events only, a more accurate and interesting comparison is between the cascade and through-going muons analyses, which, however, still results in a quite evident spectral difference.

Notice that these two datasets give ``complementary'' indications, as cascades come from the whole sky and are due to all flavors of neutrinos, with likely a preference for electron and tau neutrinos at high energies, while through-going muons are due to muon neutrinos coming from the Northern sky.

\begin{figure}
 \centering
 \includegraphics[width=\textwidth]{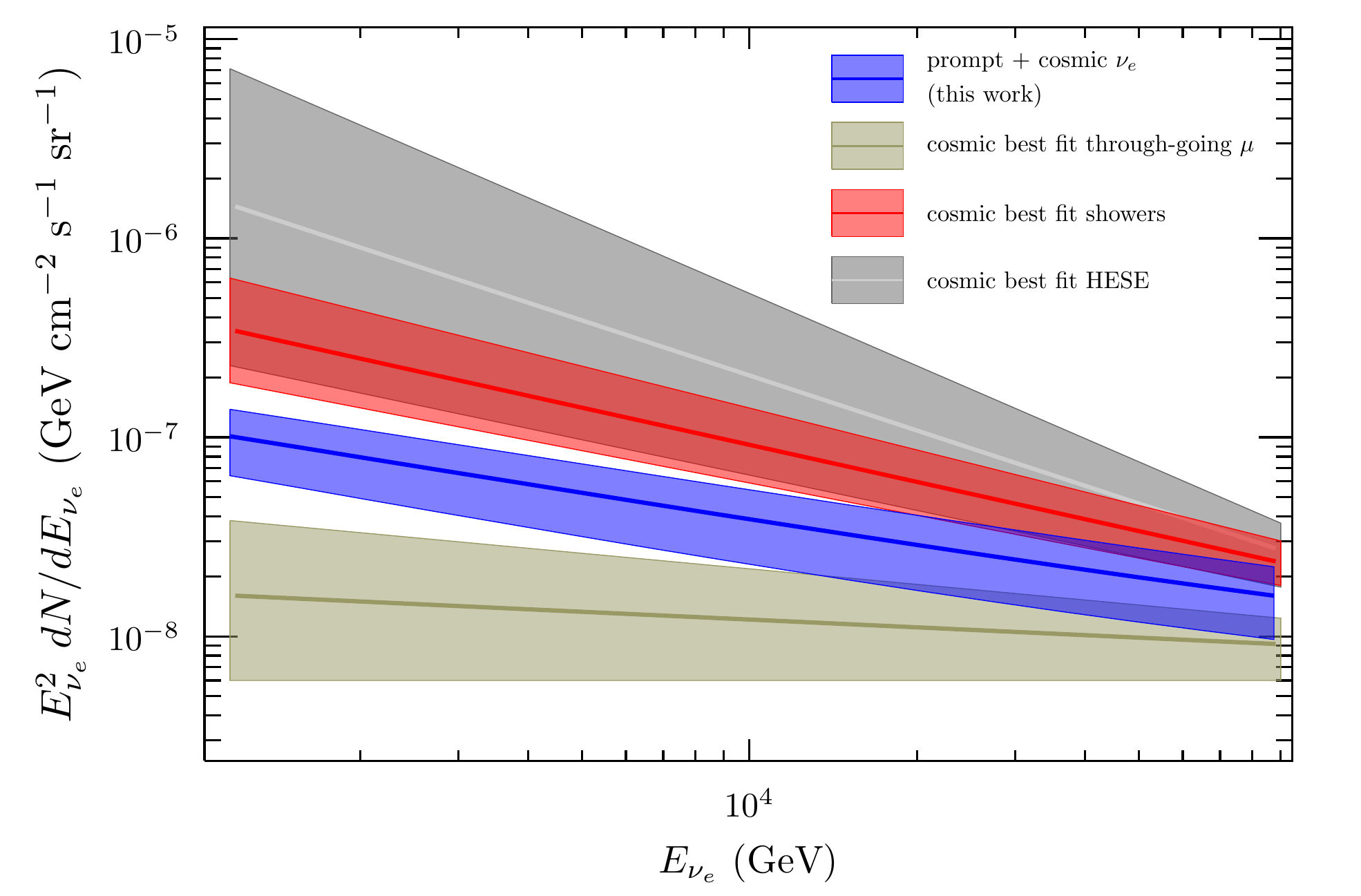} 
 \caption{The sum of the prompt and cosmic components of the $\nu_e$ spectrum as computed in this work (blue band) confronted to the astrophysical neutrino fluxes resulting from fitting the cascade sample \cite{niederhausen_high_2018} (red band), from the 6-year through-going muon analysis \cite{throughgoing} (brown band), and from the 6-year HESE analysis \cite{kopper_observation_2018} (grey band).
 The grey and red bands are experimental results, in that they come from analyses by IceCube, while the blue one is theoretical and the brown one is a low-energy extrapolation.}
 \label{fig:promptastro_nue}
\end{figure}

\begin{figure}
 \centering
 \includegraphics[width=\textwidth]{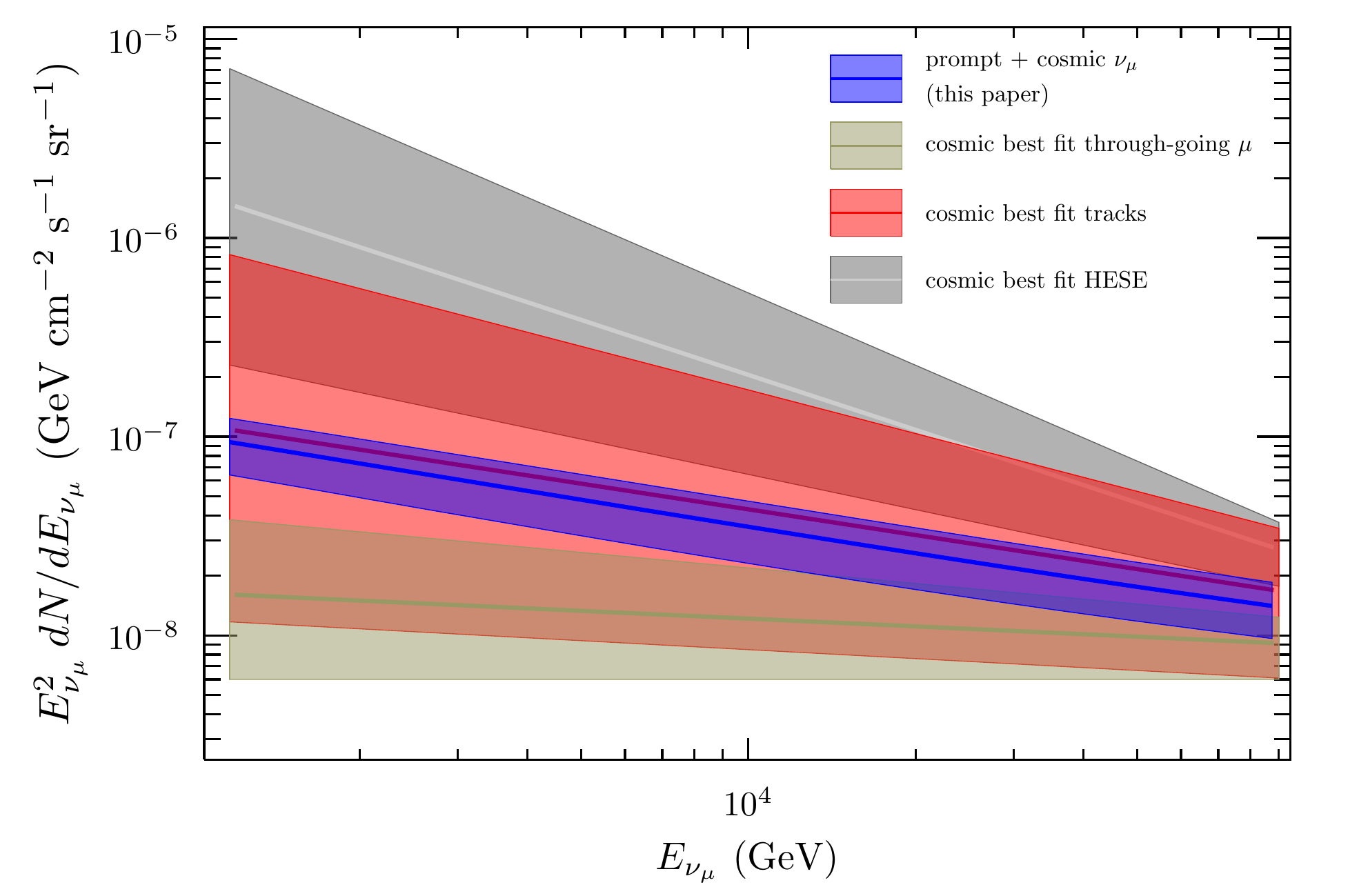} 
 \caption{The sum of the prompt and cosmic components of the $\nu_\mu$ spectrum as computed in this work (blue band) confronted to the astrophysical neutrino fluxes resulting from fitting the starting tracks sample \cite{niederhausen_high_2018} (red band), from the 6-year through-going muon analysis \cite{throughgoing} (brown band), and from the 6-year HESE analysis \cite{kopper_observation_2018} (grey band).
 The grey and red bands are experimental results, in that they come from analyses by IceCube, while the blue one is theoretical and the brown one is a low-energy extrapolation.}
 \label{fig:promptastro_numu}
\end{figure}

From table \ref{tab:showers} and figure \ref{fig:parent} a very interesting feature emerges: the number of prompt and cosmic signals in the cascade dataset with $E_\nu^\text{th} \simeq \SI{1}{TeV}$ are very similar to each other, both in the expected rate of events and in the range of energy in which they contribute. 
Taking into account also that prompt and cosmic neutrinos are expected to be isotropically distributed in the sky, it appears then difficult\footnote{It is possible when the cascade accompanying prompt neutrinos can be tagged.} to disentangle the prompt component from the astrophysical one between \SI{1}{TeV} and \SI{100}{TeV} in the cascade (and thus, HESE) dataset. 
This is consistent with the idea that the sum of the ($\sim E^{-2.7}$) prompt and the ($\sim E^{-2.13}$) astrophysical components could produce the $\sim E^{-2.62}$ spectrum obtained in the cascade analysis.
This is demonstrated in figure \ref{fig:promptastro_nue}, where we show
\begin{itemize}
\item the best-fit cosmic neutrino spectrum from the HESE analysis \cite{kopper_observation_2018} 
\item the best-fit cosmic neutrino spectrum from the cascade analysis \cite{niederhausen_high_2018} 
\item the best-fit cosmic neutrino spectrum from the through-going muons analysis \cite{throughgoing}
\item the sum of the prompt and cosmic components as computed with the model defined in section \ref{sec:expected_fluxes}.
\end{itemize}
We show only the $\nu_e$ flavor contribution in figure \ref{fig:promptastro_nue} because it is the most relevant for the cascade dataset due to astrophysical neutrinos, as seen in table \ref{tab:showers}. 
While the resemblance of the spectral shape due to the sum of cosmic and prompt neutrino fluxes and the astrophysical best fits from \cite{niederhausen_high_2018,kopper_observation_2018} is not perfect, the tension between the analyses below \SI{10}{TeV} seems somewhat alleviated.
The spectral shape obviously does not change summing over the flavors, so that this result holds true, but the sum of our prompt and cosmic components would be slightly smaller than three times the best fits from \cite{niederhausen_high_2018,kopper_observation_2018}.
From this figure, it is evident that:
\begin{itemize}
\item the spectra resulting from the through-going muons and the HESE (and cascade) analyses are not compatible at low energy, which gives rise to the ``spectral anomaly'' of the cosmic neutrino spectrum; 
\item the theoretical expectations for the sum of prompt and cosmic neutrinos, instead, 
agrees within $1\sigma$ with the best-fit cosmic neutrino flux from the cascade analysis.
\end{itemize}
We conclude that the cause of the alleged spectral anomaly can be attributed to two factors;
\begin{enumerate}
\item a prompt component does contribute to the cascade dataset in the low-energy region, $\lesssim \SI{100}{TeV}$. 
\item a part of the HESE dataset is subject to background contamination due to tracks especially at the lowest energies 
$\lesssim \SI{10}{TeV}$. 
\end{enumerate}
At this point of the discussion, it is useful to bear in mind a couple of important considerations: 
\begin{enumerate}
\item[(i)] the effectiveness of a veto system, based on the presence of muons accompanying the events with a contained vertex \cite{veto}, is better in the energy range relevant for the search for a cosmic neutrino signal - namely, above several tens of TeV - rather than in the region of lower energies, which is most relevant for the search for prompt neutrinos instead;
\item[(ii)] the same analysis that has obtained the cascade dataset \cite{niederhausen_high_2018} has been able to extract also a sample that is highly enriched in muons instead. Its power law description requires a slope of $2.43^{+0.28}_{-0.30}$, whose error is 3-4 times larger than for the cascade dataset and therefore is much less informative\footnote{This is not surprising since the conventional atmospheric component, to be subtracted, is much larger for muon neutrinos  - see figure \ref{fig:convastro} - and therefore, it is harder to identify new components in the track dataset.}.
\end{enumerate}
We performed nonetheless the same exercise considering muon neutrinos, i.e.~comparing the starting tracks best fit and the sum of prompt and astrophysical muon neutrino flux computed in this work.
As can be understood from figure \ref{fig:promptastro_numu}, no information can be extracted due to the very large uncertainties on the starting tracks sample, which could be due to difficulties in excluding atmospheric contamination.

There are other (non-exclusive) explanations of the low-energy discrepancy between the cosmic neutrino spectrum as resulting from HESE analysis and from the through-going muons analysis: a part of the low-energy soft spectrum of HESE could be due to neutrinos from the Galactic plane, and in this case
one would expect a peculiar angular distribution \cite{spurio_constraints_2014,troitsky_search_2015,neronov_evidence_2016,palladino_extragalactic_2016,pagliaroli_expectations_2016}
(which to date is not seen \cite{albert_joint_2018}); 
a priori, there could be other sources of extraterrestrial neutrinos which could cause such effect
(but this would be at odds with the null search of prompt neutrinos \cite{throughgoing}). 
Note that our proposal, concerning the role of prompt neutrinos in the cascade dataset, does not require the inclusion of hypothetical physical ingredients, and in this sense can be considered minimal.

\section{Conclusions} 
 
In this work we obtained predictions for the components of the high-energy ($\gtrsim \SI{100}{GeV}$) neutrino spectrum.

We defined a primary cosmic-ray flux, which has been tuned to the AMS-02 \cite{ams02p,ams02he} and KASCADE-Grande \cite{KG} proton and helium data, to compute the conventional and prompt components of the atmospheric neutrino spectrum.

Regarding cosmic neutrinos, we embraced hadronic collisions in dense sources, like star-forming/starburst Galaxies, as the production mechanism of astrophysical neutrinos, and combined the flux from \cite{waxman} and the best fit from \cite{throughgoing} to obtain a precise phenomenological muon neutrino flux.
We accurately described neutrino oscillations updating the formalism introduced in \cite{villaviss} (Appendix \ref{appendix2}) with the recent oscillation parameters from \cite{bari}, and predicted the astrophysical electron and tau neutrino fluxes from the muon neutrino one.

We computed the theoretical 1$\sigma$ bands for all these fluxes and compared them to the available data, obtaining a satisfactory agreement, as can be seen in figures \ref{fig:numu_components} and \ref{fig:nue_components}.

Our results are compatible with the fact that no prompt component has been seen in the through-going muon analysis, which regards muon neutrinos only.
We thus proceeded to investigate the feasibility of a measurement of prompt neutrinos from the most relevant dataset, i.e.~the muon-neutrino depleted cascade dataset.
The yearly rate of cascades in IceCube due to all flavors and components of the neutrino spectrum has been computed (table \ref{tab:showers}).
In this framework, the extraction of the prompt-neutrino signal from the cascade dataset seems difficult, but possible.

We also addressed the so called ``spectral anomaly'' \cite{palladino_icecube_2016,andrea_palladino_compatibility_2017} of the low-energy ($\lesssim \SI{100}{TeV}$) part of the cosmic neutrino spectrum arising from the different IceCube analyses \cite{throughgoing,niederhausen_high_2018,kopper_observation_2018}.
Upon visual comparison (figure \ref{fig:promptastro_nue}), it is apparent how the spectral anomaly could be attributed to the presence of prompt neutrinos, which would be the reason of the discrepancy between the $\sim E^{-2.13}$ spectrum from the through-going muon analysis and the $\sim E^{-2.48}$ one from the cascade analysis. 

We believe that great care should be taken when trying to disentangling atmospheric, and especially prompt, neutrinos from astrophysical ones between \SI{1}{TeV} and \SI{200}{TeV}, and that a global analysis, adopting the same theoretical models for all datasets would be the best way to extract informations from the data.

\appendix
\begin{appendices}

\section{Neutrino oscillation/survival probabilities and their distributions}
\label{appendix2}
In order to account for neutrino oscillations, we followed the methodology of \cite{palla}, which is based on the observation that the oscillation matrix $P_{\ell\ell'}$ is symmetric under $\ell \leftrightarrow \ell'$:
\[P_{\ell\ell'}=\sum_{i=1}^3 |U_{\ell i}^2| |U_{\ell' i}^2| \qquad \ell,\ell' = e,\mu,\tau \] 
and thus they worked out a ``natural'' parametrization with just three independent parameters, named $P_0$, $P_1$ and $P_2$.
$P_{\ell\ell'}$ is then parametrized as follows:
\[P_{\ell\ell'}=\begin{pmatrix} 
1/3+ 2P_0 &1/3-P_0+P_1 &1/3-P_0-P_1 \\[1mm]
&1/3+P_0/2-P_1+P_2 &1/3+P_0/2-P_2 \\[1mm]
&&1/3+P_0/2+P_1+P_2
\end{pmatrix}\]
with
\begin{align*}
P_0 &=\frac{1}{2} \left[(1-\epsilon)^2 \left( 1-\frac{\sin^2 (2\theta_{12})}{2}\right)+\epsilon^2 -\frac{1}{3} \right]\\[2mm]
P_1 &=\frac{1-\epsilon}{2}\left(\gamma\cos2\theta_{12}+\beta\frac{1-3\epsilon}{2} \right)\\[2mm]
P_2 &=\frac{1}{2}\left[\gamma^2 + \frac{3}{4}\beta^2 (1-\epsilon)^2 \right]
\end{align*}
and
\[\epsilon = \sin^2\theta_{13} \quad 
\alpha = \sin\theta_{13}\cos\delta\sin2\theta_{12}\sin2\theta_{23} \quad 
\beta = \cos2\theta_{23} \quad 
\gamma = \alpha - \frac{\beta}{2}\cos2\theta_{12}(1+\epsilon)\]

As for the oscillation parameters, we considered their distributions as reported in \cite{bari} for both normal ordering (NO) and inverted ordering (IO).
We performed Monte Carlo extractions of these parameters to compute the mode and the 68\% confidence intervals for the parameters $P_0$, $P_1$, $P_2$, which are reported in table \ref{tab:naturalp}.

\begin{table}
\centering
\begin{tabular}{cccc}
\midrule
 ordering &$P_0$ &$P_1$ &$P_2$ \\ 
 \midrule
NO
&\multirow{2}{*}{$0.113\pm0.006$} &$0.0345^{+0.010}_{-0.012}$&$0.0075^{+0.0045}_{-0.0038}$\\[2mm]
IO&&$0.0285^{+0.010}_{-0.057}$&$0.008^{+0.005}_{-0.006}$\\
\midrule
\end{tabular}
\caption{The natural parameters obtained from Monte Carlo sampling the oscillation parameters from \cite{bari} according to their distribution.}
\label{tab:naturalp}
\end{table}

The errors are reported with the symbol $\pm$ whenever the distribution of the relevant parameter is gaussian-like, while they are reported asymmetrically (even though they could be equal) for non-gaussian distributions. 
All these considerations apply also to table \ref{tab:oscp}, in which we report the values and confidence intervals for the oscillation/survival probabilities $P_{\ell\ell'}$, which we obtained from the same Monte Carlo extractions used to compute the distribution of $P_0$, $P_1$, and $P_2$.

\begin{figure}
 \centering
 \includegraphics[width=\textwidth,keepaspectratio]{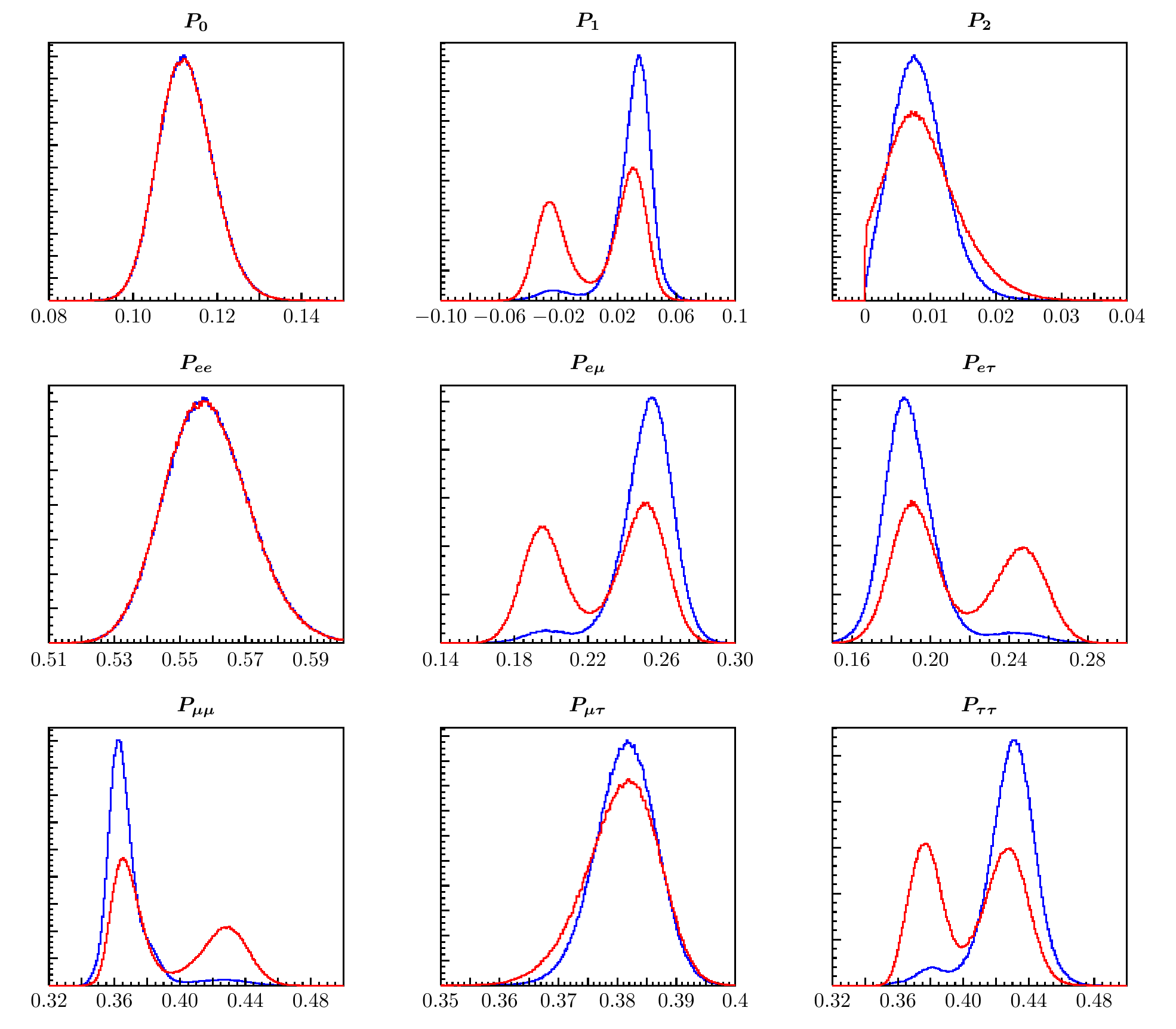}
 \caption{The distributions of the natural parameters $P_0$, $P_1$ and $P_2$ and of the oscillation/survival probabilities $P_{\ell\ell'}$ obtained with our Monte Carlo extractions using the oscillation parameters from \cite{bari}. The blue lines refer to the normal ordering of the oscillation parameters, while the red ones to the inverted ordering.}
 \label{fig:osc_ba}
\end{figure}

\begin{table}
\resizebox{\textwidth}{!}{\begin{tabular}{ccccccc}
\midrule
ordering&$P_{ee}$ &$P_{e\mu}$ &$P_{e\tau}$ &$P_{\mu\mu}$ &$P_{\mu\tau}$ &$P_{\tau\tau}$ \\ 
 \midrule
NO&\multirow{2}{*}{$0.56\pm0.01$}&$0.25^{+0.02}_{-0.01}$&$0.19^{+0.01}_{-0.02}$&$0.37^{+0.01}_{-0.02}$&$0.381\pm 0.005$&$0.43^{+0.02}_{-0.01}$\\[2mm]
IO&&$0.23^{+0.04}_{-0.03}$&$0.21^{+0.04}_{-0.04}$&$0.39^{+0.04}_{-0.04}$&$0.381\pm0.006$&$0.40^{+0.03}_{-0.03}$\\
\midrule
\end{tabular}}
\caption{The survival/oscillation probabilities obtained from Monte Carlo sampling the oscillation parameters from \cite{bari} according to their distribution.}
\label{tab:oscp}
\end{table}
In figure \ref{fig:osc_ba} we show the distributions of $P_i$ and $P_{\ell\ell'}$.
We can conclude that there is compatibility with the results obtained in \cite{palla}.

\section{The kernels linking neutrinos to photons}
\label{appendix1}
Due to the change over time of the oscillation parameters measurements with time, we find it useful to provide the reader with a resume of the 
kernels to obtain neutrinos from photons, as in eq.~\eqref{eq:appendila}; in fact the oscillated kernels, which we denote with a $\sim$, are easily obtained from the non-oscillated ones like so:
\[\tilde K_{\nu_\ell} = \sum_{\ell'= e,\mu}  P_{\ell\ell'} K_{\nu_{\ell'}} \qquad \ell = e,\mu,\tau\]
The generic form of the non-oscillated kernels is the following: 
\[
K_{\nu_\ell}(x) =\alpha_\pi\delta\left(x-(1-r_\pi)\right)+\alpha_K\delta\left(x-(1-r_K)\right)+
\begin{dcases}
x^2 (\beta_0+\beta_1 x) &x\leq r_K\\[2mm]
\sum_{n=0}^3 \chi_n x^n &r_K < x < r_\pi\\[2mm]
(1-x)^2 (\delta_0+\delta_1 x) &x\geq r_\pi
\end{dcases}\]
and their parameters are listed in table \ref{tab:kernpars}, while $r_i = (m_\mu/m_i)^2 $.
\begin{table}
\centering
\includegraphics[width=\textwidth,keepaspectratio]{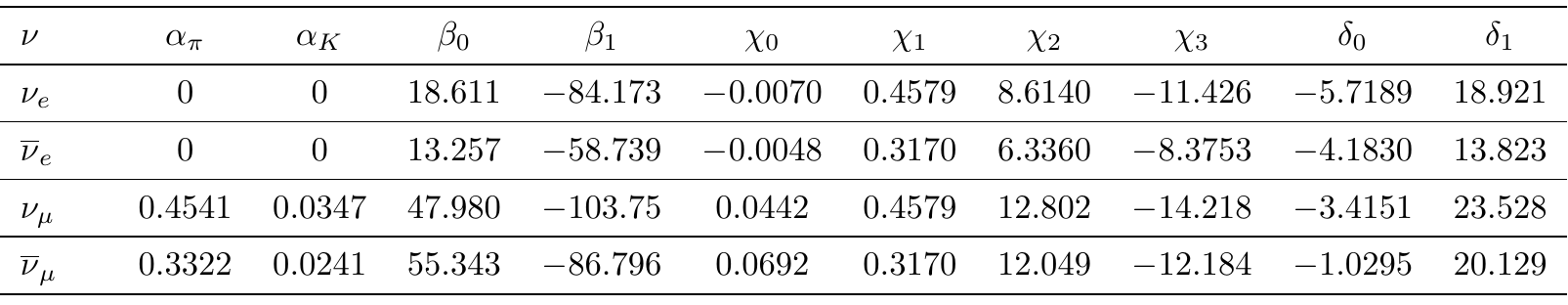}
\caption{The parameters for the non-oscillated kernels.}
\label{tab:kernpars}
\end{table}

\end{appendices}

\acknowledgments

The authors are thankful to Andrea Palladino, Maria Vittoria Garzelli, Paolo Lipari, Spencer Klein and the anonymous referee for valuable discussions and feedback on the text.
This work was partially supported by the research grant number 2017W4HA7S ``NAT-NET: Neutrino and Astroparticle Theory Network'' under the program PRIN 2017 funded by the Italian Ministero dell'Istruzione, dell'Universit\`a e della Ricerca (MIUR). 

\newpage
\addcontentsline{toc}{section}{References}
\bibliographystyle{jhep}
\bibliography{refs}

\begin{thebibliography}{10}
\expandafter\ifx\csname url\endcsname\relax
  \def\url#1{\texttt{#1}}\fi
\expandafter\ifx\csname urlprefix\endcsname\relax\def\urlprefix{URL }\fi
\expandafter\ifx\csname href\endcsname\relax
  \def\href#1#2{#2} \def\path#1{#1}\fi

\bibitem{aartsen_observation_2014}
\relax The {IceCube}~Collaboration,
  \href{http://arxiv.org/abs/1405.5303}{Observation of {High}-{Energy}
  {Astrophysical} {Neutrinos} in {Three} {Years} of {IceCube} {Data}}, Physical
  Review Letters 113~(10), arXiv: 1405.5303.
\newblock \href {http://dx.doi.org/10.1103/PhysRevLett.113.101101}
  {\path{doi:10.1103/PhysRevLett.113.101101}}.
\newline\urlprefix\url{http://arxiv.org/abs/1405.5303}

\bibitem{throughgoing}
\relax The IceCube~Collaboration,
  \href{http://arxiv.org/abs/1607.08006}{Observation and characterization of a
  cosmic muon neutrino flux from the northern hemisphere using six years of
  {IceCube} data}, The Astrophysical Journal 833~(1) (2016) 3.
\newblock \href {http://arxiv.org/abs/1607.08006} {\path{arXiv:1607.08006}},
  \href {http://dx.doi.org/10.3847/0004-637X/833/1/3}
  {\path{doi:10.3847/0004-637X/833/1/3}}.
\newline\urlprefix\url{http://arxiv.org/abs/1607.08006}

\bibitem{ic_nue_atm}
\relax The {IceCube}~Collaboration,
  \href{http://arxiv.org/abs/1504.03753}{Measurement of the atmospheric $\nu_e$
  spectrum with {IceCube}}, Physical Review D 91~(12).
\newblock \href {http://arxiv.org/abs/1504.03753} {\path{arXiv:1504.03753}},
  \href {http://dx.doi.org/10.1103/PhysRevD.91.122004}
  {\path{doi:10.1103/PhysRevD.91.122004}}.
\newline\urlprefix\url{http://arxiv.org/abs/1504.03753}

\bibitem{ic_numu_atm}
\relax The {IceCube}~Collaboration,
  \href{http://arxiv.org/abs/1409.4535}{Development of a general analysis and
  unfolding scheme and its application to measure the energy spectrum of
  atmospheric neutrinos with {IceCube}}, The European Physical Journal C
  75~(3).
\newblock \href {http://arxiv.org/abs/1409.4535} {\path{arXiv:1409.4535}},
  \href {http://dx.doi.org/10.1140/epjc/s10052-015-3330-z}
  {\path{doi:10.1140/epjc/s10052-015-3330-z}}.
\newline\urlprefix\url{http://arxiv.org/abs/1409.4535}

\bibitem{txs1}
\relax The~IceCube, Fermi-LAT, Magic, Agile, Asas-Sn, HAWK, HESS, Integral,
  Kanata, Kiso, Kapteyn, L.~Telescope, Subaru, Swift/NuSTAR, Veritas, V.-.
  Teams,
  \href{http://science.sciencemag.org/content/early/2018/07/11/science.aat1378}{Multimessenger
  observations of a flaring blazar coincident with high-energy neutrino
  {IceCube}-170922a}, Science\href {http://dx.doi.org/10.1126/science.aat1378}
  {\path{doi:10.1126/science.aat1378}}.
\newline\urlprefix\url{http://science.sciencemag.org/content/early/2018/07/11/science.aat1378}

\bibitem{ic_icrc17}
\relax The {IceCube}~Collaboration, \href{http://arxiv.org/abs/1710.01191}{The
  {IceCube} neutrino observatory - contributions to {ICRC} 2017 part {II}:
  Properties of the atmospheric and astrophysical neutrino flux},
  {arXiv}:1710.01191 [astro-ph]\href {http://arxiv.org/abs/1710.01191}
  {\path{arXiv:1710.01191}}.
\newline\urlprefix\url{http://arxiv.org/abs/1710.01191}

\bibitem{palladino_icecube_2016}
A.~Palladino, M.~Spurio, F.~Vissani, \href{http://arxiv.org/abs/1610.07015}{On
  the {IceCube} spectral anomaly}, Journal of Cosmology and Astroparticle
  Physics 2016~(12) (2016) 045--045, arXiv: 1610.07015.
\newblock \href {http://dx.doi.org/10.1088/1475-7516/2016/12/045}
  {\path{doi:10.1088/1475-7516/2016/12/045}}.
\newline\urlprefix\url{http://arxiv.org/abs/1610.07015}

\bibitem{andrea_palladino_compatibility_2017}
A.~Palladino, C.~Mascaretti, F.~Vissani,
  \href{https://epjc.epj.org/articles/epjc/abs/2017/10/10052_2017_Article_5273/10052_2017_Article_5273.html}{On
  the compatibility of the {IceCube} results with a universal neutrino
  spectrum}, The European Physical Journal C 77~(10) (2017) 684.
\newblock \href {http://dx.doi.org/10.1140/epjc/s10052-017-5273-z}
  {\path{doi:10.1140/epjc/s10052-017-5273-z}}.
\newline\urlprefix\url{https://epjc.epj.org/articles/epjc/abs/2017/10/10052_2017_Article_5273/10052_2017_Article_5273.html}

\bibitem{palladino_icecube_2018}
A.~Palladino, A.~Fedynitch, R.~W. Rasmussen, A.~M. Taylor,
  \href{https://arxiv.org/abs/1812.04685}{{IceCube} {Neutrinos} from
  {Hadronically} {Powered} {Gamma}-{Ray} {Galaxies}}, arXiv.
\newline\urlprefix\url{https://arxiv.org/abs/1812.04685}

\bibitem{ams02p}
{The AMS-02 Collaboration}, {Precision Measurement of the Proton Flux in
  Primary Cosmic Rays from Rigidity 1 GV to 1.8 TV with the Alpha Magnetic
  Spectrometer on the International Space Station}, Physical Review Letters
  114~(17) (2015) 171103.
\newblock \href {http://dx.doi.org/10.1103/PhysRevLett.114.171103}
  {\path{doi:10.1103/PhysRevLett.114.171103}}.

\bibitem{ams02he}
{The AMS-02 Collaboration}, {Precision Measurement of the Helium Flux in
  Primary Cosmic Rays of Rigidities 1.9 GV to 3 TV with the Alpha Magnetic
  Spectrometer on the International Space Station}, Physical Review Letters
  115~(21) (2015) 211101.
\newblock \href {http://dx.doi.org/10.1103/PhysRevLett.115.211101}
  {\path{doi:10.1103/PhysRevLett.115.211101}}.

\bibitem{KG}
{The KASCADE-Grande Collaboration}, Ankle-like feature in the energy spectrum
  of light elements of cosmic rays observed with {KASCADE}-grande, Physical
  Review D 87~(8).
\newblock \href {http://arxiv.org/abs/1304.7114} {\path{arXiv:1304.7114}},
  \href {http://dx.doi.org/10.1103/PhysRevD.87.081101}
  {\path{doi:10.1103/PhysRevD.87.081101}}.

\bibitem{fedynitch_state---art_2018}
A.~Fedynitch, H.~Dembinski, R.~Engel, T.~K. Gaisser, F.~Riehn, T.~Stanev,
  \href{https://pos.sissa.it/301/1019/}{A state-of-the-art calculation of
  atmospheric lepton fluxes}, in: Proceedings of 35th {International} {Cosmic}
  {Ray} {Conference} — {PoS}({ICRC}2017), Vol. 301, SISSA Medialab, 2018, p.
  1019.
\newblock \href {http://dx.doi.org/10.22323/1.301.1019}
  {\path{doi:10.22323/1.301.1019}}.
\newline\urlprefix\url{https://pos.sissa.it/301/1019/}

\bibitem{TIG}
M.~Thunman, G.~Ingelman, P.~Gondolo,
  \href{http://arxiv.org/abs/hep-ph/9505417}{Charm production and high energy
  atmospheric muon and neutrino fluxes}, Astroparticle Physics 5~(3) (1996)
  309--332.
\newblock \href {http://arxiv.org/abs/hep-ph/9505417}
  {\path{arXiv:hep-ph/9505417}}, \href
  {http://dx.doi.org/10.1016/0927-6505(96)00033-3}
  {\path{doi:10.1016/0927-6505(96)00033-3}}.
\newline\urlprefix\url{http://arxiv.org/abs/hep-ph/9505417}

\bibitem{sarcevic}
R.~Enberg, M.~H. Reno, I.~Sarcevic,
  \href{http://arxiv.org/abs/0806.0418}{Prompt neutrino fluxes from atmospheric
  charm}, Physical Review D 78~(4).
\newblock \href {http://arxiv.org/abs/0806.0418} {\path{arXiv:0806.0418}},
  \href {http://dx.doi.org/10.1103/PhysRevD.78.043005}
  {\path{doi:10.1103/PhysRevD.78.043005}}.
\newline\urlprefix\url{http://arxiv.org/abs/0806.0418}

\bibitem{honda}
M.~Honda, T.~Kajita, K.~Kasahara, S.~Midorikawa, T.~Sanuki,
  \href{http://arxiv.org/abs/astro-ph/0611418}{Calculation of atmospheric
  neutrino flux using the interaction model calibrated with atmospheric muon
  data}, Physical Review D 75~(4).
\newblock \href {http://arxiv.org/abs/astro-ph/0611418}
  {\path{arXiv:astro-ph/0611418}}, \href
  {http://dx.doi.org/10.1103/PhysRevD.75.043006}
  {\path{doi:10.1103/PhysRevD.75.043006}}.
\newline\urlprefix\url{http://arxiv.org/abs/astro-ph/0611418}

\bibitem{rottoli}
R.~Gauld, J.~Rojo, L.~Rottoli, S.~Sarkar, J.~Talbert,
  \href{http://arxiv.org/abs/1511.06346}{The prompt atmospheric neutrino flux
  in the light of {LHCb}}, Journal of High Energy Physics 2016~(2).
\newblock \href {http://arxiv.org/abs/1511.06346} {\path{arXiv:1511.06346}},
  \href {http://dx.doi.org/10.1007/JHEP02(2016)130}
  {\path{doi:10.1007/JHEP02(2016)130}}.
\newline\urlprefix\url{http://arxiv.org/abs/1511.06346}

\bibitem{garzelli}
M.~Benzke, M.~V. Garzelli, B.~A. Kniehl, G.~Kramer, S.~Moch, G.~Sigl, Prompt
  neutrinos from atmospheric charm in the general-mass variable-flavor-number
  scheme, Journal of High Energy Physics 2017~(12).
\newblock \href {http://arxiv.org/abs/1705.10386} {\path{arXiv:1705.10386}},
  \href {http://dx.doi.org/10.1007/JHEP12(2017)021}
  {\path{doi:10.1007/JHEP12(2017)021}}.

\bibitem{berss}
A.~Bhattacharya, R.~Enberg, Y.~S. Jeong, C.~S. Kim, M.~H. Reno, I.~Sarcevic,
  A.~Stasto, \href{http://arxiv.org/abs/1607.00193}{Prompt atmospheric neutrino
  fluxes: perturbative {QCD} models and nuclear effects}, Journal of High
  Energy Physics 2016~(11).
\newblock \href {http://arxiv.org/abs/1607.00193} {\path{arXiv:1607.00193}},
  \href {http://dx.doi.org/10.1007/JHEP11(2016)167}
  {\path{doi:10.1007/JHEP11(2016)167}}.
\newline\urlprefix\url{http://arxiv.org/abs/1607.00193}

\bibitem{laha_ic_2017}
R.~Laha, S.~J. Brodsky, \href{http://arxiv.org/abs/1607.08240}{{IC} at {IC}:
  {IceCube} can constrain the intrinsic charm of the proton}, Physical Review D
  96~(12) (2017) 123002, arXiv: 1607.08240.
\newblock \href {http://dx.doi.org/10.1103/PhysRevD.96.123002}
  {\path{doi:10.1103/PhysRevD.96.123002}}.
\newline\urlprefix\url{http://arxiv.org/abs/1607.08240}

\bibitem{bhattacharya_forward_2018}
A.~Bhattacharya, J.~R. Cudell, \href{http://arxiv.org/abs/1808.00293}{Forward
  charm-production models and prompt neutrinos at {IceCube}}, Journal of High
  Energy Physics 2018~(11) (2018) 150, arXiv: 1808.00293.
\newblock \href {http://dx.doi.org/10.1007/JHEP11(2018)150}
  {\path{doi:10.1007/JHEP11(2018)150}}.
\newline\urlprefix\url{http://arxiv.org/abs/1808.00293}

\bibitem{giannini_intrinsic_2018}
A.~V. Giannini, V.~P. Goncalves, F.~S. Navarra,
  \href{http://arxiv.org/abs/1803.01728}{Intrinsic charm contribution to the
  prompt atmospheric neutrino flux}, Physical Review D 98~(1) (2018) 014012,
  arXiv: 1803.01728.
\newblock \href {http://dx.doi.org/10.1103/PhysRevD.98.014012}
  {\path{doi:10.1103/PhysRevD.98.014012}}.
\newline\urlprefix\url{http://arxiv.org/abs/1803.01728}

\bibitem{MCEQ}
A.~{Fedynitch}, R.~{Engel}, T.~K. {Gaisser}, F.~{Riehn}, T.~{Stanev},
  {Calculation of conventional and prompt lepton fluxes at very high energy},
  in: European Physical Journal Web of Conferences, Vol.~99 of European
  Physical Journal Web of Conferences, 2015, p. 08001.
\newblock \href {http://arxiv.org/abs/1503.00544} {\path{arXiv:1503.00544}},
  \href {http://dx.doi.org/10.1051/epjconf/20159908001}
  {\path{doi:10.1051/epjconf/20159908001}}.

\bibitem{tanabashi_review_2018}
\relax M.~{Tanabashi}~et al., Review of {Particle} {Physics}, Phys.Rev. D98~(3)
  (2018) 030001.
\newblock \href {http://dx.doi.org/10.1103/PhysRevD.98.030001}
  {\path{doi:10.1103/PhysRevD.98.030001}}.

\bibitem{mbe_2019}
C.~Mascaretti, P.~Blasi, C.~Evoli,
  \href{http://arxiv.org/abs/1906.05197}{Atmospheric neutrinos and the knee of
  the {Cosmic} {Ray} spectrum}, Astroparticle Physics 114 (2019) 22--29, arXiv:
  1906.05197.
\newblock \href {http://dx.doi.org/10.1016/j.astropartphys.2019.06.002}
  {\path{doi:10.1016/j.astropartphys.2019.06.002}}.
\newline\urlprefix\url{http://arxiv.org/abs/1906.05197}

\bibitem{syb2.3}
F.~{Riehn}, R.~{Engel}, A.~{Fedynitch}, T.~K. {Gaisser}, T.~{Stanev}, {A new
  version of the event generator Sibyll}, ArXiv e-prints\href
  {http://arxiv.org/abs/1510.00568} {\path{arXiv:1510.00568}}.

\bibitem{nrlmsise}
J.~M. Picone, A.~E. Hedin, D.~P. Drob, A.~C. Aikin,
  \href{https://agupubs.onlinelibrary.wiley.com/doi/abs/10.1029/2002JA009430}{{NRLMSISE}-00
  empirical model of the atmosphere: {Statistical} comparisons and scientific
  issues}, Journal of Geophysical Research: Space Physics 107~(A12) (2002) SIA
  15--1--SIA 15--16.
\newblock \href {http://dx.doi.org/10.1029/2002JA009430}
  {\path{doi:10.1029/2002JA009430}}.
\newline\urlprefix\url{https://agupubs.onlinelibrary.wiley.com/doi/abs/10.1029/2002JA009430}

\bibitem{waxman}
A.~Loeb, E.~Waxman, \href{http://arxiv.org/abs/astro-ph/0601695}{The cumulative
  bakground of high-energy neutrinos from starburst galaxies}, Journal of
  Cosmology and Astroparticle Physics 2006~(5) (2006) 003--003.
\newblock \href {http://arxiv.org/abs/astro-ph/0601695}
  {\path{arXiv:astro-ph/0601695}}, \href
  {http://dx.doi.org/10.1088/1475-7516/2006/05/003}
  {\path{doi:10.1088/1475-7516/2006/05/003}}.
\newline\urlprefix\url{http://arxiv.org/abs/astro-ph/0601695}

\bibitem{villaviss}
F.~L. Villante, F.~Vissani,
  \href{https://link.aps.org/doi/10.1103/PhysRevD.78.103007}{How precisely can
  neutrino emission from supernova remnants be constrained by gamma ray
  observations?}, Physical Review D 78~(10) (2008) 103007.
\newblock \href {http://dx.doi.org/10.1103/PhysRevD.78.103007}
  {\path{doi:10.1103/PhysRevD.78.103007}}.
\newline\urlprefix\url{https://link.aps.org/doi/10.1103/PhysRevD.78.103007}

\bibitem{bari}
F.~Capozzi, E.~Di~Valentino, E.~Lisi, A.~Marrone, A.~Melchiorri, A.~Palazzo,
  \href{https://link.aps.org/doi/10.1103/PhysRevD.95.096014}{Global constraints
  on absolute neutrino masses and their ordering}, Physical Review D 95~(9)
  (2017) 096014.
\newblock \href {http://dx.doi.org/10.1103/PhysRevD.95.096014}
  {\path{doi:10.1103/PhysRevD.95.096014}}.
\newline\urlprefix\url{https://link.aps.org/doi/10.1103/PhysRevD.95.096014}

\bibitem{niederhausen_high_2018}
\relax {The IceCube Collaboration},
  \href{http://arxiv.org/abs/1808.07629}{Measurements using the inelasticity
  distribution of multi-{TeV} neutrino interactions in {IceCube}}, Physical
  Review D 99~(3) (2019) 032004, arXiv: 1808.07629.
\newblock \href {http://dx.doi.org/10.1103/PhysRevD.99.032004}
  {\path{doi:10.1103/PhysRevD.99.032004}}.
\newline\urlprefix\url{http://arxiv.org/abs/1808.07629}

\bibitem{denton_invisible_2018}
P.~B. Denton, I.~Tamborra, \href{http://arxiv.org/abs/1805.05950}{Invisible
  {Neutrino} {Decay} {Resolves} {IceCube}'s {Track} and {Cascade} {Tension}},
  Physical Review Letters 121~(12) (2018) 121802, arXiv: 1805.05950.
\newblock \href {http://dx.doi.org/10.1103/PhysRevLett.121.121802}
  {\path{doi:10.1103/PhysRevLett.121.121802}}.
\newline\urlprefix\url{http://arxiv.org/abs/1805.05950}

\bibitem{palladino_multi-component_2018}
A.~Palladino, W.~Winter, \href{http://arxiv.org/abs/1801.07277}{A
  {Multi}-{Component} {Model} for the {Observed} {Astrophysical} {Neutrinos}},
  Astronomy \& Astrophysics 615 (2018) A168, arXiv: 1801.07277.
\newblock \href {http://dx.doi.org/10.1051/0004-6361/201832731}
  {\path{doi:10.1051/0004-6361/201832731}}.
\newline\urlprefix\url{http://arxiv.org/abs/1801.07277}

\bibitem{sui_combined_2018}
Y.~Sui, P.~S.~B. Dev, \href{http://arxiv.org/abs/1804.04919}{A {Combined}
  {Astrophysical} and {Dark} {Matter} {Interpretation} of the {IceCube} {HESE}
  and {Throughgoing} {Muon} {Events}}, Journal of Cosmology and Astroparticle
  Physics 2018~(07) (2018) 020--020, arXiv: 1804.04919.
\newblock \href {http://dx.doi.org/10.1088/1475-7516/2018/07/020}
  {\path{doi:10.1088/1475-7516/2018/07/020}}.
\newline\urlprefix\url{http://arxiv.org/abs/1804.04919}

\bibitem{kopper_observation_2018}
C.~Kopper, on~behalf of~the IceCube~Collaboration,
  \href{https://pos.sissa.it/cgi-bin/reader/contribution.cgi?id=PoS(ICRC2017)981}{Observation
  of {Astrophysical} {Neutrinos} in {Six} {Years} of {IceCube} {Data}}, in:
  Proceedings of 35th {International} {Cosmic} {Ray} {Conference} —
  {PoS}({ICRC}2017), Vol. 301, SISSA Medialab, 2018, p. 981.
\newblock \href {http://dx.doi.org/10.22323/1.301.0981}
  {\path{doi:10.22323/1.301.0981}}.
\newline\urlprefix\url{https://pos.sissa.it/cgi-bin/reader/contribution.cgi?id=PoS(ICRC2017)981}

\bibitem{veto}
S.~Schönert, T.~K. Gaisser, E.~Resconi, O.~Schulz,
  \href{https://arxiv.org/abs/0812.4308v1}{Vetoing atmospheric neutrinos in a
  high energy neutrino telescope}, Physical Review D\href
  {http://dx.doi.org/10.1103/PhysRevD.79.043009}
  {\path{doi:10.1103/PhysRevD.79.043009}}.
\newline\urlprefix\url{https://arxiv.org/abs/0812.4308v1}

\bibitem{spurio_constraints_2014}
M.~Spurio, \href{http://arxiv.org/abs/1409.4552}{Constraints to a {Galactic}
  {Component} of the {Ice} {Cube} cosmic neutrino flux from {ANTARES}},
  Physical Review D 90~(10) (2014) 103004, arXiv: 1409.4552.
\newblock \href {http://dx.doi.org/10.1103/PhysRevD.90.103004}
  {\path{doi:10.1103/PhysRevD.90.103004}}.
\newline\urlprefix\url{http://arxiv.org/abs/1409.4552}

\bibitem{troitsky_search_2015}
S.~Troitsky, \href{http://arxiv.org/abs/1511.01708}{Search for {Galactic} disk
  and halo components in the arrival directions of high-energy astrophysical
  neutrinos}, JETP Letters 102~(12) (2015) 785--788, arXiv: 1511.01708.
\newblock \href {http://dx.doi.org/10.1134/S0021364015240133}
  {\path{doi:10.1134/S0021364015240133}}.
\newline\urlprefix\url{http://arxiv.org/abs/1511.01708}

\bibitem{neronov_evidence_2016}
A.~Neronov, D.~V. Semikoz, \href{http://arxiv.org/abs/1509.03522}{Evidence for
  the {Galactic} contribution to the {IceCube} astrophysical neutrino flux},
  Astroparticle Physics 75 (2016) 60--63, arXiv: 1509.03522.
\newblock \href {http://dx.doi.org/10.1016/j.astropartphys.2015.11.002}
  {\path{doi:10.1016/j.astropartphys.2015.11.002}}.
\newline\urlprefix\url{http://arxiv.org/abs/1509.03522}

\bibitem{palladino_extragalactic_2016}
A.~Palladino, F.~Vissani, \href{http://arxiv.org/abs/1601.06678}{Extragalactic
  plus {Galactic} model for {IceCube} neutrino events}, The Astrophysical
  Journal 826~(2) (2016) 185, arXiv: 1601.06678.
\newblock \href {http://dx.doi.org/10.3847/0004-637X/826/2/185}
  {\path{doi:10.3847/0004-637X/826/2/185}}.
\newline\urlprefix\url{http://arxiv.org/abs/1601.06678}

\bibitem{pagliaroli_expectations_2016}
G.~Pagliaroli, C.~Evoli, F.~L. Villante,
  \href{http://arxiv.org/abs/1606.04489}{Expectations for high energy diffuse
  galactic neutrinos for different cosmic ray distributions}, Journal of
  Cosmology and Astroparticle Physics 2016~(11) (2016) 004--004, arXiv:
  1606.04489.
\newblock \href {http://dx.doi.org/10.1088/1475-7516/2016/11/004}
  {\path{doi:10.1088/1475-7516/2016/11/004}}.
\newline\urlprefix\url{http://arxiv.org/abs/1606.04489}

\bibitem{albert_joint_2018}
\relax The {IceCube and ANTARES}~Collaborations, {D.~Gaggero}, {D.~Grasso},
  \href{http://arxiv.org/abs/1808.03531}{Joint constraints on {Galactic}
  diffuse neutrino emission from {ANTARES} and {IceCube}}, arXiv:1808.03531
  [astro-ph]ArXiv: 1808.03531.
\newline\urlprefix\url{http://arxiv.org/abs/1808.03531}

\bibitem{palla}
A.~Palladino, F.~Vissani,
  \href{https://link.springer.com/article/10.1140/epjc/s10052-015-3664-6}{The
  natural parameterization of cosmic neutrino oscillations}, The European
  Physical Journal C 75~(9) (2015) 433.
\newblock \href {http://dx.doi.org/10.1140/epjc/s10052-015-3664-6}
  {\path{doi:10.1140/epjc/s10052-015-3664-6}}.
\newline\urlprefix\url{https://link.springer.com/article/10.1140/epjc/s10052-015-3664-6}

\end{thebibliography}

\end{document}